\documentclass[preprint,aps,tightenlines,nofootinbib]{revtex4}
\usepackage{graphicx}
\usepackage{bm}
\usepackage{epsfig}
\usepackage{amssymb}
\usepackage{amsmath}

\newcommand{\ba}{\begin{eqnarray}}
\newcommand{\ea}{\end{eqnarray}}
\newcommand{\bea}{\begin{eqnarray}}
\newcommand{\eea}{\end{eqnarray}}

\newcommand{\be}{\begin{equation}}
\newcommand{\ee}{\end{equation}}
\newcommand{\slsh}[1]{\not\!{#1}}
\newcommand{\nln}{\nonumber}

\begin{document}

\hfill$\vcenter{
\hbox{\bf MADPH-06-1458} }$ 

\hfill$\vcenter{
\hbox{\bf SU-4252-825} }$


\title{
\vspace*{0.5cm}
Probing the Randall-Sundrum geometric origin of flavor with lepton flavor violation}




\author{
Kaustubh Agashe$^a$,
Andrew E. Blechman$^b$,
Frank Petriello$^c$\\
}

\affiliation{
$^a$Department~of~Physics,~Syracuse University,~Syracuse,~NY~13244
\\ and \\
School of Natural Sciences, Institute for Advanced Study, Princeton, NJ 08540
\\
$^b$Department of Physics, The Johns Hopkins University, Baltimore, MD 21218\\
$^c$Department of Physics, University of Wisconsin, Madison, WI  53706}

\begin{abstract}

The ``anarchic'' Randall-Sundrum model of flavor is a low energy
solution to both the electroweak hierarchy and flavor problems.  Such models have a warped, compact extra
dimension with the standard model fermions and gauge bosons living in the
bulk, and the Higgs living on or near the TeV brane. In this paper we
consider bounds on these models set by lepton flavor violation constraints.  We find that loop-induced decays of the form
$l \to l^{'}\gamma$ are ultraviolet sensitive and uncalculable when the
Higgs field is localized on a four-dimensional brane; this drawback
does not occur when the Higgs field propagates in the full five-dimensional space-time.
We find constraints at the few TeV level throughout
the natural range of parameters, arising from $\mu-e$ conversion in the presence of nuclei, rare $\mu$ decays,
and rare $\tau$ decays.  A ``tension" exists between loop-induced dipole decays such as $\mu \to e\gamma$ and tree-level
processes such as $\mu-e$ conversion; they have opposite dependences on the five-dimensional Yukawa couplings, making
it difficult to decouple flavor-violating effects.  
We emphasize the importance of the future experiments MEG and PRIME.  
These experiments will definitively test the Randall-Sundrum
geometric origin of hierarchies in the lepton sector at the TeV-scale.

\end{abstract}

\maketitle


\section{Introduction}

The Standard Model (SM) of particle physics is a remarkably successful description 
of nature.  However, it contains several unsatisfactory features.  
In particular, there are many hierarchies built into the model that
have no {\it \`{a} priori} explanation.  The most famous of these is the 
huge separation between the electroweak and Planck scales.  There have been many proposed solutions to this problem.  One
possibility is the Randall-Sundrum scenario (RS)~\cite{RS1}.  In this model, our 
four-dimensional space-time is embedded into a five-dimensional anti de-Sitter space.  The 
extra ``warped'' fifth dimension is compactified on an orbifold.  This 
space-time is described by the metric
\begin{equation}
ds^2=e^{-2kr_c\phi}\eta_{\mu\nu}dx^\mu dx^\nu - r_c^2 d\phi^2,
\end{equation}
where $-\pi\le\phi\le\pi$.  
Three-branes are placed at the orbifold fixed points $\phi=0$ and $\phi=\pi$
(and its reflection at $\phi = - \pi$).  
The brane at $\phi=0$ is called the Planck or ultraviolet (UV) brane, while the brane at $\phi=\pi$ is called
the TeV or infrared (IR) brane.  For sizes of the fifth dimension $kr_c \sim 11-12$, the TeV scale 
is obtained from the fundamental Planck scale via an exponential warping induced by the anti de-Sitter geometry: 
$M_{{\rm TeV}} = M_{{\rm pl}} e^{-k\pi r_c}$.  It was shown that this setup can be naturally 
stabilized~\cite{Goldberger:1999uk}.  The original model placed all SM fields on the IR brane.

This scenario does not explain all unnatural parameters in the SM.  The fermion Yukawa couplings, except 
for the top quark coupling, are small and hierarchical.  The minimal RS model offers no 
solution to this flavor hierarchy problem.  In addition, the flavor sector in the RS model is 
sensitive to ultraviolet physics, and requires a cut-off of roughly $10^3 \, {\rm TeV}$ to avoid
dangerous flavor-changing neutral currents.  This is problematic, as the only cut-off 
available is the electroweak scale.

One solution to this problem is to permit some or all of the SM fields to propagate in the 
full $5D$ space
\cite{Davoudiasl:1999tf, GN, GP}.  
The only requirement for solving the gauge hierarchy problem is to have the 
Higgs field localized near the IR brane.  This immediately presents a solution to the flavor hierarchy problem \cite{GN, GP}, 
since the Yukawa couplings of the Higgs field to the fermions become dependent on the position
of the fermion fields relative to the IR brane.  By placing fermions at different positions in the 
$5D$ bulk, a hierarchy in the effective $4D$ Yukawa couplings can be generated 
even with anarchic $\mathcal{O}(1)$ $5D$ couplings.  These ``anarchic'' RS models 
set all diagonal and off-diagonal Yukawa couplings to $\mathcal{O}(1)$.  In addition, 
allowing fermions to propagate in the bulk suppresses the operators leading to dangerous 
flavor changing neutral currents~\cite{GP, Huber:2000ie}.
Some collider \cite{DHRoffwall} and flavor physics \cite{Huber:2003tu, Burdman:2003nt, Khalil:2004qk, APS, 
Moreau:2006np} phenomenology of these models has been 
considered previously.

Additional work is needed to make this scenario fully realistic.  It was shown that the simplest 
formulation leads to large violations of the custodial symmetry in the SM~\cite{HPR}.  There are 
two known solutions to this problem.  The first extends the bulk gauge symmetry to $SU(2)_L\times SU(2)_R$; when 
broken by boundary conditions, a bulk custodial $SU(2)$ symmetry is preserved~\cite{ADMS}.  The second model 
introduces large brane kinetic terms to suppress precision electroweak constraints~\cite{BKT}.  Both solutions 
allow for the masses of the first Kaluza-Klein (KK) excitations to be as low as 3 TeV, generating 
interesting phenomenology which may be observable at the upcoming Large Hadron Collider (LHC).

In this paper we probe the anarchic RS scenario by examining its effects on lepton flavor-violating 
observables.  We study here a minimalistic model; we assume the SM gauge group, KK masses of a few 
TeV or larger, and an anarchic $5D$ Yukawa structure.  We allow the Higgs boson to propagate in the full 
$5D$ space, which encompasses features found in several recent models~\cite{Agashe:2004rs,DLR}.  
Specific theories such as those mentioned above 
with a left-right symmetric bulk or large brane kinetic terms will predict slightly different 
effects than we find here, but we believe that our analysis captures the most important effects.  We note that 
the flavor violation we study here is completely independent of neutrino physics parameters.  We subject the 
anarchic RS picture to a complete set of experimental constraints: the rare $\mu$ decays $\mu \to e\gamma$ and 
$\mu^{\pm} \to e^+e^-e^{\pm}$, the rare $\tau$ decays $\tau \to \{e,\mu\}\gamma$ and tri-lepton decay modes, and $\mu-e$ conversion 
in the presence of nuclei.  We find constraints on the KK scale of a few TeV throughout parameter space.  Interestingly, 
there is a ``tension'' between dipole operator decays such as $l \to l^{'}\gamma$ and the remaining 
processes.  They have different dependences on the $5D$ Yukawa parameters, leading to strong 
constraints throughout parameter space.  We also find that when the Higgs field is localized on the TeV brane, the 
dipole decays $l \to l^{'}\gamma$ are UV sensitive and uncalculable in the RS theory.  This does not occur when the 
Higgs boson can propagate in the full $5D$ space-time.  We emphasize the important role played by several future 
experiments: MEG~\cite{Signorelli:2003vw}, which will improve the 
constraints on $\mu \to e\gamma$ by two orders of magnitude; PRIME~\cite{PRIMEref}, which will strengthen the bounds 
on $\mu-e$ conversion by several orders of magnitude; super-$B$ factories, which will improve the 
bounds on rare $\tau$ decays by an order of magnitude.  Measurements from these three experiments 
will definitively test the anarchic RS picture.

We briefly compare our work to previous papers on lepton flavor violation in the RS framework.  
Reference~\cite{Kitano:2000wr} studied lepton flavor violation in a scenario where only 
a right-handed neutrino propagates in the full $5D$ spacetime.  The studies 
in~\cite{Huber:2003tu,Moreau:2006np} allowed all SM fermions and gauge bosons to propagate in the bulk.
Reference~\cite{Huber:2003tu} did not incorporate custodial isospin, and therefore 
considered KK masses of 10 TeV, while the paper~\cite{Moreau:2006np} considered a 
model with structure in the $5D$ masses and Yukawa couplings.  None of these 
studies considered a bulk Higgs field.  They also did not address the UV sensitivity 
of dipole decays in the brane Higgs field scenario, nor did they discuss the 
tension between tree-level and loop-induced processes.  We also present a more detailed study 
of future experimental prospects than previous analyses.

This paper is organized as follows.  In Section 2 we present our notation and describe the model.  We discuss 
in Section 3 the $\mu-e$ conversion and tri-lepton decay processes, which are mediated by tree-level 
gauge boson mixing.  We discuss the loop-induced decays $l \to l^{'}\gamma$ in Section 4.  In Section 5 we 
present our Monte Carlo scan over the anarchic RS parameter space.  We summarize and conclude in Section 6.

\section{Notation and Conventions}

In this section we present our notation and describe the model we consider.  The basic action is 
\begin{equation}
S=\int d^4xd\phi\sqrt{G}[\mathcal{L}_{{\rm gauge}}+\mathcal{L}_{{\rm fermion}}+\mathcal{L}_{{\rm Higgs}}].
\end{equation}
The Lagrangian for gauge fields in the bulk, $\mathcal{L}_{{\rm gauge}}$, has been studied in~\cite{Davoudiasl:1999tf}.  
$\mathcal{L}_{{\rm fermion}}$ was presented in~\cite{GN, GP, DHRoffwall} 
using an IR brane Higgs boson; we will review the 
relevant formulae and discuss the transition to a bulk Higgs below.  Our setup of the bulk Higgs field will 
follow the discussion in~\cite{DLR}.

\subsection{Brane Higgs field}

We begin by considering the case of the Higgs field localized on the IR brane.  The Lagrangian in this case 
is
\begin{equation}
\mathcal{L}_{\rm Higgs}=\left[D_\mu H(D^\mu H)^\dagger-V(H)-\mathcal{L}_{\rm Yukawa}\right] 
\Big[ \delta(\phi-\pi) + \delta (\phi + \pi ) \Big],
\end{equation}
where $D_\mu$ is the covariant derivative. 
$\mathcal{L}_{\rm Yukawa}$ describes the Yukawa
interactions with the fermions.  The Lagrangian for bulk fermions was
derived in~\cite{GN, GP, DHRoffwall}; it takes the form
\begin{equation}
\mathcal{L}_{\rm fermion}=i\bar{\Psi}E^M_A\Gamma^AD_M\Psi
-{\rm sgn}(\phi)kc_\Psi\bar{\Psi}\Psi.
\end{equation}
where $E^M_A$ is the inverse vielbein.  This Lagrangian admits zero-mode solutions.  The $c_\Psi$ parameters indicate where in the fifth 
dimension 
the zero-mode fermions are localized: either near the TeV brane ($c<1/2$) or near the Planck brane ($c>1/2$).  
The $4D$ Yukawa couplings of these fermions are exponentially sensitive to the $c_\Psi$ parameters.
We perform the KK decomposition of the fermion 
field by splitting it into chiral components, $\Psi=\Psi_L+\Psi_R$, yielding
\begin{equation}
\Psi_{L,R}(x,\phi)=\sum_n\frac{e^{2kr_c|\phi|}}{\sqrt{r_c}}\psi^{(n)}_{L,R}(x)
f_{L,R}^{(n)}(\phi;c)
\label{KKexp}.
\end{equation}
The $c$ dependence becomes part of the KK wavefunction 
$f_{L,R}^{(n)}(\phi;c)$; explicit formulas for these wavefunctions can be found 
in~\cite{GN, GP, DHRoffwall}.

The SM contains two types of fermions, corresponding to singlets ($S$) and
doublets ($D$) under $SU(2)_L$.  In the SM, we require that the $S$ fermions are
right-handed while the $D$ fermions are left-handed.  However, in five
dimensions we must have both chiralities.  To get a chiral zero-mode sector we 
use the orbifold parity of RS models.  In particular, we choose $(S_R,D_L)$
to be even under the orbifold parity (Neuman boundary conditions) and
$(S_L,D_R)$ to be odd (Dirichlet boundary conditions).  The odd fields
will not have zero modes, and the even zero modes will correspond to the SM
fermions.  We now group these fermions and their first KK modes into the 
vectors
\begin{eqnarray}
\Psi_L^I&=&(D_L^{i(0)},D_L^{i(1)},S_L^{i(1)}), \nonumber \\
\Psi_R^I&=&(S_R^{i(0)},S_R^{i(1)},D_R^{i(1)}),
\label{KKfstates}
\end{eqnarray}
where $i$ is a flavor index ($i=e,\mu,\tau$) and $I=1...9$.  We will show in a later section 
that higher KK modes have a negligible effect on our results.

The fundamental $5D$ Yukawa interaction is
\begin{equation}
\mathcal{L}_{\rm Yukawa}=\frac{\lambda_{5D}^{ij}}{k}\bar{D}_L^{i} H S_{R}^{j}.
\end{equation}
Using the vectors in Eq.~\ref{KKfstates} and substituting in the KK expansion of 
Eq.~\ref{KKexp} yields
\begin{equation}
\mathcal{L}_{\rm Yukawa}=\frac{\Lambda_{IJ}}{k}\bar{\Psi}_L^{I}H\Psi_R^J +{\rm h.c.},
\end{equation}
where
\begin{equation}
\Lambda=\left(\begin{array}{ccc}
\lambda_{4D}    & \lambda_{4D}F_R    & 0 \\
F_L\lambda_{4D} & F_L\lambda_{4D}F_R & 0 \\
0 & 0 & 0  \end{array}\right).
\label{yukawa}
\end{equation}
Each internal block is a $3\times 3$ matrix, with
\begin{equation}
F_{L,R}\equiv\left(\begin{array}{ccc}
f_{e_{L,R}}^{(1)}/f_{e_{L,R}}^{(0)} & 0 & 0 \\
0 & f_{\mu_{L,R}}^{(1)}/f_{\mu_{L,R}}^{(0)} & 0 \\ 
0 & 0 & f_{\tau_{L,R}}^{(1)}/f_{\tau_{L,R}}^{(0)}
\end{array}\right)
\end{equation}
These should be evaluated on the TeV brane, since that is where the Higgs is localized.  We find
\begin{equation}
\lambda_{4D}^{ij}=\frac{\epsilon}{kr_c}f^{(0)}_if^{(0)}_j\lambda_{5D}^{ij}
=\sqrt{\frac{(1-2c_i)(1-2c_j)}{(\epsilon^{1-2c_i}-1)(\epsilon^{1-2c_j}-1)}}
\epsilon^{1-(c_i+c_j)}\times\lambda_{5D}^{ij},
\label{4dyuk}
\end{equation}
where $\epsilon=e^{\pi kr_c}$ and there is no sum over $i,j$.  
It is straightforward to write down the mass matrix for the fermions:
\begin{equation}
\mathcal{M}=\left(\begin{array}{ccc}
M_0     & M_0F_R    & 0         \\
F_LM_0  & F_LM_0F_R & M_{KK}    \\
0       & M_{KK}    & 0
\end{array}\right),
\end{equation}
where $M_0^{ij}=\frac{v}{\sqrt{2}}\lambda_{4D}^{ij}$ is the zero mode mass matrix.  
$M_{KK}$ is a diagonal matrix that contains the KK masses.  $M_0$ is
not diagonal.  We can diagonalize this zero mode mass matrix in the usual
way, by  constructing a biunitary transformation $(U_L,U_R)$ so that $M_D=U_LM_0U_R^\dagger$ is
diagonal.  We can embed this rotation into the full matrix above by multiplying
on the left by diag($U_L,1,1$) and on the right by diag($U_R^\dagger,1,1$). 
This gives
\begin{equation}
\mathcal{M}=\left(\begin{array}{ccc}
M_D        		    & \frac{v}{\sqrt{2}}\Delta_R  & 0       \\
\frac{v}{\sqrt{2}}\Delta_L  & \Delta_1	  		  & M_{KK}  \\
0          		    & M_{KK}     		  & 0
\end{array}\right).
\label{mass}
\end{equation}
We have set$\frac{v}{\sqrt{2}}\Delta_R=U_LM_0F_R=M_DU_RF_R$ and
$\frac{v}{\sqrt{2}}\Delta_L=F_LM_0U_R^\dagger=F_LU_L^{\dagger}M_D$.  
A factor of $\frac{v}{\sqrt{2}}$ was extracted to make it easier to match to the 
Yukawa matrix.  Notice that the middle entry can also be written in terms of the
diagonal zero-mode matrix: $\Delta_1=F_LM_0F_R=F_LU_L^{\dagger}M_DU_RF_R$.  From now on, we will 
use this expression.  To find the Yukawa matrix $\Lambda$ in this
basis, we just divide Eq.~\ref{mass} by $\frac{v}{\sqrt{2}}$ and 
set $M_{KK}=0$.  We note that this implies we are considering the exchange of a complex Higgs boson, 
which is equivalent to the exchange of 
the physical Higgs boson
and the longitudinal component of the $Z$.
The diagonalization of this mass matrix is discussed in the Appendix.

\subsection{Bulk Higgs field}

We now discuss the changes that occur when we allow the Higgs to propagate in the full $5D$ space.  
A new coupling of the Higgs boson to the fermion KK states exists: $H\bar{S}_L^{(1)}D_R^{(1)}+$h.c.  
This is not present in the brane Higgs case because the $S_L$ and
$D_R$ wavefunctions vanish identically on the TeV brane due to the Dirichlet boundary
conditions.  The fermion mass matrix becomes
\begin{equation}
\mathcal{M}=\left(\begin{array}{ccc}
M_D        		    & \frac{v}{\sqrt{2}}\Delta_R   & 0           \\
\frac{v}{\sqrt{2}}\Delta_L  & \Delta_1		           & M_{KK}      \\
0  		            & M_{KK}   		           & \Delta_2
\end{array}\right).
\label{massbulk}
\end{equation}
$\Delta_{L,R,1}$ are not the same as in the brane case; they now include 
overlap integrals of the $KK$ and zero-mode fermion wavefunctions with the Higgs
wavefunction.  $\Delta_2$ represents the wavefunction overlaps between the first KK modes of the right-handed doublet 
and left-handed singlet leptons; the explicit expressions as well as the details of diagonalizing $\mathcal{M}$ can be found in the Appendix.  
We note that all of the $\Delta$ are proportional to the $4D$ Yukawa couplings.  

Our discussion of the bulk Higgs field will follow the presentation in~\cite{DLR}.  The $5D$ profile 
for the Higgs vev is
\begin{equation}
\chi_H (\phi)=N_H e^{2\sigma}J_{\nu}\left( i x_T 
e^{kr_c(\phi-\pi)}\right).
\label{prof1}
\end{equation}
Here, $x_T$ is the solution of a root equation giving the tachyonic mass, $N_H$ is a normalization factor, $\sigma=kr_c\phi$, 
and $\nu$ is the index of the solution.  We will simplify this further for our 
discussion by using the asymptotic expansion of the Bessel function for large index, $J_{\nu}(z) \sim
z^{\nu}$.  
Using this expansion 
gives the following normalized profile:
\begin{equation}
\chi_H (\phi)= \sqrt{\frac{kr_c(1+\nu)}{e^{2(1+\nu)kr\pi}-1}} e^{(2+\nu)\sigma}.
\label{prof2}
\end{equation}
This satisfies the constraint 
\begin{equation}
1=2\int_{0}^{\pi} d\phi e^{-2\sigma} \chi_H^2 (\phi),
\end{equation}
where the factor of 2 comes from the $[-\pi,0]$ integration.  In our analysis we will 
vary the index $\nu$, without worrying about its dependence on the model
parameters in~\cite{DLR}.  This 
also makes a connection with the $A_5$ composite Higgs models in~\cite{Agashe:2004rs}, which is
approximately realized in this framework as $\nu=0$.  We can also make a direct comparison to
the TeV brane Higgs scenario, which is realized by $\nu\rightarrow\infty$.

\begin{figure}[htb]
\centerline{
\psfig{figure=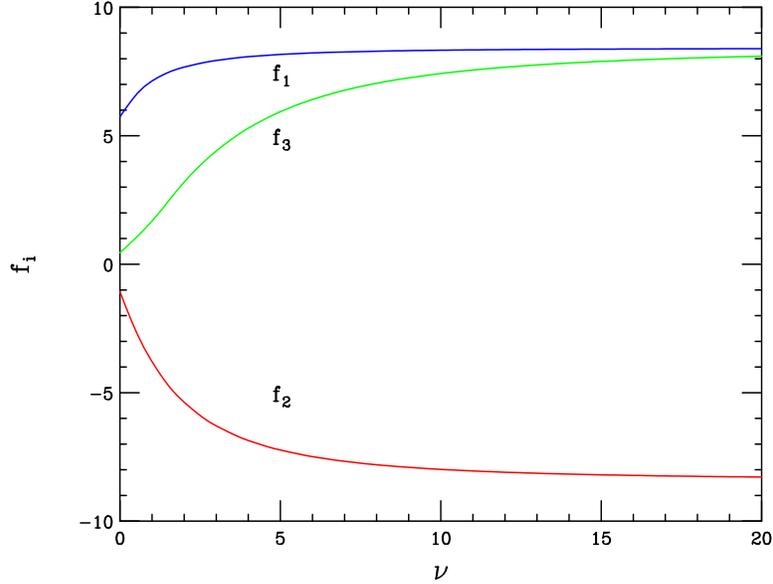,height=10.2cm,width=7.8cm,angle=90}}
\vspace{-0.2cm}
\caption{$f_i$, the off-diagonal elements of the gauge boson mass matrix that describe the mixing of the 
zero-mode with the $i$-th KK-mode.}
\label{fplot}
\end{figure}

We will now study the effect of the bulk Higgs field on the gauge boson sector.  We begin with the action
\begin{equation}
S_{gauge}=\int d^5x \sqrt{-G} \, G^{MN}\left(D_M H\right)^{\dag} D_N H.
\end{equation}
Performing a standard KK decomposition, and expanding $H=v \chi_H /\sqrt{2r_c}$, we arrive at the mass matrix
\begin{equation}
\frac{m_{0}^2}{2} \sum_{m,n=0} a_{mn} A_{\mu}^{(m)} A^{\mu (n)},
\end{equation}
with
\begin{equation}
a_{mn} = 4\pi \int_{0}^{\pi} d\phi \, e^{-2\sigma} \chi_{H}^{2} \chi^{(m)} \chi^{(n)}.
\label{gaugemass}
\end{equation}
The $\chi^{(n)}$ are the usual gauge wave-functions, which can be found in~\cite{Davoudiasl:1999tf}.  
We note that $\chi^{(0)} =1/\sqrt{2\pi}$.  We show in
Fig.~\ref{fplot} the elements 
$f_i=a_{0i}$ of this mixing matrix.  The expectation is that as $\nu \rightarrow \infty$, these should approach the brane Higgs 
values of $(-1)^{i+1} \sqrt{2\pi kr_c} \approx \pm 8.42$, assuming the value $kr_c=11.27$; this is indeed what occurs.  
\noindent
%

We must now study the fermion sector, particularly what form the $4D$ Yukawa couplings take 
in terms of the $5D$ values.  We begin with the action
\begin{equation}
S_{ffH} = \frac{\lambda_{5D}\sqrt{1+\nu}}{\sqrt{k}} \int d^5x \, \sqrt{-G} H^{\dag} \psi_D \psi_S^c.
\end{equation}
where the $\nu$-dependent prefactor is included to reproduce the correct $4D$ Yukawa coupling as
$\nu\rightarrow\infty$.  The zero-mode fermion wave-function is $e^{2\sigma}f^{(0)}/\sqrt{r_c}$, where 
\begin{equation}
f^{(0)} = \sqrt{\frac{kr_c(1-2c)}{2\left(e^{(1-2c)kr_c\pi}-1\right)}} e^{-c\sigma}.
\label{zerowf}
\end{equation}
Inserting this into the action, and expanding $H$ as before, we find the following expression for the 4D Yukawa:
\begin{eqnarray}
\lambda_{4D}&=& \frac{\lambda_{5D}\left(1-2c\right)}{e^{(1-2c)kr_c\pi}-1} \left[\sqrt{kr_c(1+\nu)}\int_{0}^{\pi} d\phi \, 
  \chi_H e^{-2c\sigma} \right]  \nonumber \\
 &=&  \frac{\lambda_{5D}\left(1-2c\right)}{e^{(1-2c)kr_c\pi}-1} \left[\frac{1+\nu}{\sqrt{e^{2(1+\nu)kr_c\pi}-1}} \frac{e^{(2+\nu-2c)kr_c\pi}
-1}{2+\nu-2c}\right].
\label{bulklambda}
\end{eqnarray}
For simplicity, we have only presented the diagonal Yukawa coupling.  To reproduce the brane Higgs diagonal Yukawa coupling in Eq.~\ref{4dyuk}, 
the bracketed integral should reduce to $e^{(1-2c)kr_c\pi}$ as $\nu \rightarrow \infty$.  It is simple to check that this occurs.

\subsection{The anarchic RS parameters}

We discuss here the parameters of the anarchic RS model and give their natural values.  We first 
note that Eq.~\ref{4dyuk} relates the diagonal $5D$ Yukawa couplings to the fermion $c$ parameters through 
the measured fermion masses.  The off-diagonal entries are removed after diagonalization with 
$U_{L,R}$.  The preferred size of the Yukawa couplings can be determined by demanding consistency with 
$Z \to b\bar{b}$ measurements and by the size of the top quark mass; this yields $\lambda_{5D} \approx 2$~\cite{ADMS}.  
We assume three couplings ($Y_e,Y_\mu,Y_\tau$) of this approximate magnitude.  The size of these 
couplings implies $c>1/2$ for all three leptons, indicating that they are localized near 
the Planck brane.  For simplicity, we take $c_L=c_R$.  We note that this range of $c$ is the appropriate one for 
first and second generation quarks also; for the third generation, $c_{b_L}=c_{t_L}\sim
0.45$, while $c_{b_R} \sim 0.5$ and $c_{t_R}\sim 0$~\cite{ADMS}.

We can also estimate the natural sizes of the $U_{L,R}$ matrix elements.  For illustration, we 
consider here a two-family scenario; it is straightforward to extend this example to three families.  Assuming 
an anarchic RS scenario, so that all of the $\lambda_{5D}^{ij}\sim\mathcal{O}(1)$, we can use Eq.~\ref{4dyuk} 
to write the $4D$ Yukawa matrix as
\begin{equation}
\lambda_{5D}=\left(\begin{array}{cc}
Y_{11}    & Y_{12} \\
Y_{12} & Y_{22}
\end{array}\right)
\Rightarrow
\lambda_{4D}\sim\left(\begin{array}{cc}
Y_{11}f^{(0)2}_e         & Y_{12} f^{(0)}_ef^{(0)}_\mu \\
Y_{12} f^{(0)}_ef^{(0)}_\mu & Y_{22} f^{(0)2}_\mu
\end{array}\right)\label{y},
\end{equation}
where we have assumed for simplicity a symmetric $5D$ Yukawa matrix.  Assuming $\mathcal{O}(1)$ Yukawa couplings, the 
functional dependences of the fermion masses on the wave-functions are
\begin{equation}
m_e \sim f^{(0)2}_e, \,\,\, m_\mu \sim f^{(0)2}_\mu, 
\end{equation}
while the mixing matrices take the form
\begin{equation}
U\sim\left(\begin{array}{cc}
1         & -\sqrt{\frac{m_e}{m_\mu}} \\
\sqrt{\frac{m_e}{m_\mu}} & 1
\end{array}\right).
\end{equation}
We therefore find that $|U_{12}|\sim\sqrt{\frac{m_e}{m_\mu}}$.  Including the $\tau$ 
then gives $|U_{13}|\sim\sqrt{\frac{m_e}{m_\tau}}$ and $|U_{23}|\sim\sqrt{\frac{m_\mu}{m_\tau}}$.  The 
diagonal entries $|U_{ii}| \sim 1$.  We will assume mixing matrix elements of these approximate magnitudes 
in our analysis.

\subsection{Operator Matching}

We discuss in this subsection the formalism we will use to compare the RS predictions to the experimental 
measurements.  Our presentation closely follows the discussion in~\cite{Chang:2005ag}.  Tri-lepton 
decays of the form $l \to l_1\bar{l}_2l_3$ and $\mu-e$ conversion are mediated by tree-level 
mixing with heavy gauge bosons and generate four-fermion interactions, while $l \to l^{'}\gamma$ 
occurs via a loop-induced dipole operator.  We can parameterize these effects in the following 
effective Lagrangian:
\begin{eqnarray}
-\mathcal{L_{{\rm eff}}}&=&C_R(q^2)\frac{1}{2m_\mu}\bar{e}_R\sigma^{\mu\nu}F_{\mu\nu}\mu_L+
C_L(q^2)\frac{1}{2m_\mu}\bar{e}_L\sigma^{\mu\nu}F_{\mu\nu}\mu_R \nln\\
& &+\frac{4G_F}{\sqrt{2}}\left[g_3(\bar{e}_R\gamma^\mu\mu_R)(\bar{e}_R\gamma_\mu e_R)
+g_4(\bar{e}_L\gamma^\mu\mu_L)(\bar{e}_L\gamma_\mu e_L)\right. \nln\\
& &+\left.g_5(\bar{e}_R\gamma^\mu\mu_R)(\bar{e}_L\gamma_\mu e_L)
+g_6(\bar{e}_L\gamma^\mu\mu_L)(\bar{e}_R\gamma_\mu e_R)\right] + {\rm h.c.}
\label{ngeq}
\end{eqnarray}
The form factors\footnote{Note that these form factors are normalized
differently than the $A_{L,R}$ in~\cite{Chang:2005ag}: $C=-\frac{8G_Fm_\mu^2}{\sqrt{2}}A$.} $C_{L,R}(q^2)$ and the couplings $g_i$ are then 
computed in a
straightforward matching procedure.  We will discuss this computation in detail in the 
following two sections.

\section{Tri-lepton decays and $\mu-e$ conversion}

In this section and the next we study the predictions that the minimal RS model makes for lepton flavor violation.  We focus on 
processes in the muon sector, such as $\mu^- \rightarrow e^+e^-e^-$ and $\mu-e$ conversion in the presence of 
nuclei, and rare tau decays of the form $\tau \to l_1 \bar{l}_2 l_3$ currently being studied at BABAR and BELLE.  
The dipole-mediated decays will be discussed in the next section.

The dominant effects arise from flavor non-diagonal couplings of the zero-mode $Z$-boson.  Contributions from exchange of the 
Higgs boson are suppressed by small fermion masses, and we will show later that those coming from direct $KK$ exchange 
are suppressed by a large fermion wave-function factor.  There are also contributions to these processes from 
the dipole exchanges denoted by $C_{L,R}$ in Eq.~\ref{ngeq}, but these are loop-suppressed and small in the 
parameter space of interest.    We also find that KK-fermion mixing effects are sub-dominant in the parameter space 
of interest.  We derive here the relevant couplings.  
We denote the physical basis by $Z_0,Z_1$, and the gauge basis by $Z^{(0)},Z^{(1)}$.  For simplicity, we 
restrict our discussion here to the first KK level; in our analysis we include the first several modes.  After 
diagonalizing the gauge boson mass matrix, we find that these are related via 
\begin{equation}
Z^{(0)}=Z_0+f\frac{m_Z^2}{M_{KK}^{2}}Z_1, \quad Z^{(1)}=Z_1-f\frac{m_Z^2}{M_{KK}^{2}}Z_0.
\label{gaugemix}
\end{equation}
$f$ parameterizes the mixing between the zero and first KK level.  With a brane Higgs field, $f=\sqrt{2k\pi r_c}\sim\mathcal{O}(10)$.  A plot 
of $f$ for a bulk Higgs field is shown in Fig.~\ref{fplot}.  The couplings between the zero-mode 
fermions and $Z^{(1)}$ are determined by the appropriate overlap integral.  We define the ratio of these couplings 
to the SM couplings as $\alpha_e$, $\alpha_{\mu}$, and $\alpha_{\tau}$, where $g^{(1)}=\alpha g^{SM}$; the $\alpha_i$ are 
then given by
\begin{equation}
\alpha_i = 2\sqrt{2\pi}\int_{0}^{\pi} d\phi \, e^{\sigma} \chi^{(1)} [f^{(0)}_i]^2.
\label{alphadef}
\end{equation}
Since the fermion 
wave-functions are localized at different points in the bulk, the $\alpha_i$ differ, but they are all
roughly $\mathcal{O}(0.1)$ in magnitude.  We present a plot of the $\alpha_i$ in Fig.~\ref{alphaplot}.  In the fermion flavor basis, 
the matrix which describes the 
$Z^{(1)}$ couplings takes the form 
\begin{equation}
g^{SM} \left(\bar{e}_F, \bar{\mu}_F, \bar{\tau}_F\right) \slsh{Z^{(1)}} \left(\begin{array}{lll} \alpha_e & 0 & 0 \\ 0 & \alpha_{\mu} & 0 \\
  0 & 0 & \alpha_{\tau} \end{array}\right) \left(\begin{array}{l} e_F \\ \mu_F \\ \tau_F \end{array}\right).
\end{equation}
We must first rotate the fermions to the mass basis.  As was explained in the
last section, we introduce unitary matrices $U_L$, $U_R$, so that 
$L_M = U_L L_F$, $R_M = U_R R_F$, where $L_F$ denotes the left-handed flavor basis-vector, $L_M$ the left-handed mass basis-vector, etc.  The 
flavor-basis coupling matrices $C^{F}_{L,R} = g_{L,R} \, {\rm diag}(\alpha_e,\alpha_{\mu},\alpha_{\tau})$ are rotated to 
$C_{L,R}=U_{L,R} \,C^{F}_{L,R} \,U^{\dag}_{L,R}$.  The flavor-violating couplings are the off-diagonal entries of $C_{L,R}$; 
we find
\begin{eqnarray}
g^{(1)\mu e}_{L,R} &=& g_{L,R} 
\left(U^{L,R}_{11}U^{L,R*}_{21}\alpha_e+U^{L,R}_{12}U^{L,R*}_{22}\alpha_\mu+U^{L,R}_{13}U^{L,R*}_{23}\alpha_\tau\right),
 \nonumber \\
g^{(1)\tau \mu}_{L,R} &=& g_{L,R} 
\left(U^{L,R}_{21}U^{L,R*}_{31}\alpha_e+U^{L,R}_{22}U^{L,R*}_{32}\alpha_\mu+U^{L,R}_{23}U^{L,R*}_{33}\alpha_\tau\right), 
 \nonumber \\
g^{(1)\tau e}_{L,R} &=& g_{L,R} 
\left(U^{L,R}_{11}U^{L,R*}_{31}\alpha_e+U^{L,R}_{12}U^{L,R*}_{32}\alpha_\mu+U^{L,R}_{13}U^{L,R*}_{33}\alpha_\tau\right),
\end{eqnarray}
where $g_{L,R}$ are the usual SM couplings.  We can use the unitarity of $U_{L,R}$ to rewrite these as 
\begin{eqnarray}
g^{(1)\mu e}_{L,R} &=& g_{L,R}\left[U^{L,R}_{12}U^{L,R*}_{22}(\alpha_\mu-\alpha_e)+U^{L,R}_{13}U^{L,R*}_{23}(\alpha_\tau-\alpha_e)\right],
 \nonumber \\
g^{(1)\tau \mu}_{L,R} &=& 
g_{L,R}\left[U^{L,R}_{21}U^{L,R*}_{31}(\alpha_e-\alpha_\mu)+U^{L,R}_{23}U^{L,R*}_{33}(\alpha_\tau-\alpha_\mu)\right],
 \nonumber \\
g^{(1)\tau e}_{L,R} &=& g_{L,R}\left[U^{L,R}_{12}U^{L,R*}_{32}(\alpha_\mu-\alpha_e)+U^{L,R}_{13}U^{L,R*}_{33}(\alpha_\tau-\alpha_e)\right].
\label{fvcoups}
\end{eqnarray}
Using Eq.~\ref{gaugemix}, the couplings to $Z_0$ are obtained via multiplication by $-f m_{Z}^2/M_{KK}^{2}$: 
$g^{\mu e}_{L,R} = -f m_{Z}^2/M_{KK}^{2}g^{(1)\mu e}_{L,R}$, etc.  The couplings to $Z_1$ are identical to those in 
Eq.~\ref{fvcoups}, to leading order in the gauge boson mixing.

We now use these to derive the flavor-violating couplings $g_{3-6}$ of
Eq.~\ref{ngeq}:
\begin{eqnarray}
g^{\mu e}_3 &=& 2g_R \left[g_{R}^{\mu e}+\alpha_e g_{R}^{(1) \mu e}\frac{m_Z^2}{M_{KK}^{2}}\right], \nonumber \\
g^{\mu e}_4 &=& 2g_L \left[g_{L}^{\mu e}+\alpha_e g_{L}^{(1) \mu e}\frac{m_Z^2}{M_{KK}^{2}}\right], \nonumber \\
g^{\mu e}_5 &=& 2g_L \left[g_{R}^{\mu e}+\alpha_e g_{R}^{(1) \mu e}\frac{m_Z^2}{M_{KK}^{2}}\right], \nonumber \\
g^{\mu e}_6 &=& 2g_R \left[g_{L}^{\mu e}+\alpha_e g_{L}^{(1) \mu e}\frac{m_Z^2}{M_{KK}^{2}}\right].
\label{fvcoups2}
\end{eqnarray}
These are for $\mu-e$ flavor violation; similar expressions hold for $\tau-\mu$ and $\tau-e$.  The first term on each line is from 
the $Z_0$ coupling, while the second is from direct $Z_1$ exchange.  Substituting in the expressions from Eq.~\ref{fvcoups}, we 
find 
\begin{equation}
g^{\mu e}_3 = -2g_{R}^{2}\frac{m_{Z}^{2}}{M_{KK}^{2}}\left(f-\alpha_e\right)\left[U^{R}_{12}U^{R*}_{22}(\alpha_\mu-\alpha_e)
  +U^{R}_{13}U^{R*}_{23}(\alpha_\tau-\alpha_e)\right],
\label{eqexam}
\end{equation}
and similar expressions for the other couplings.  Since $f \gg |\alpha_e|$, we can neglect the direct KK exchange effect.

We will study the decays $\mu^- \rightarrow e^-e^+e^-$,  $\tau^- \rightarrow \mu^-\mu^+\mu^-$, 
$\tau^- \rightarrow e^-e^+e^-$, $\tau \rightarrow \mu^-e^+e^-$, and $\tau \rightarrow e^-\mu^+\mu^-$.  The remaining 
rare $\tau$ decays studied at BABAR and BELLE, $\tau \rightarrow e^-\mu^+e^-$ and $\tau \rightarrow \mu^-e^+\mu^-$, require 
an additional flavor-violating coupling than those above, and are therefore highly suppressed.  The relevant branching 
fractions from~\cite{Chang:2005ag} are 
\begin{eqnarray}
BR(\mu \rightarrow 3e) &=& 2\left(|g^{\mu e}_3|^2+|g^{\mu e}_4|^2\right)+|g^{\mu e}_5|^2+|g^{\mu e}_6|^2, \nonumber \\
BR(\tau \rightarrow 3\mu) &=& \left\{2\left(|g^{\tau \mu}_3|^2+|g^{\tau \mu}_4|^2\right)+|g^{\tau \mu}_5|^2+|g^{\tau \mu}_6|^2 
   \right\} BR(\tau \to e\nu\nu), \nonumber \\
BR(\tau \rightarrow 3e) &=& \left\{2\left(|g^{\tau e}_3|^2+|g^{\tau e}_4|^2\right)+|g^{\tau e}_5|^2+|g^{\tau e}_6|^2 
   \right\} BR(\tau \to e\nu\nu),\nonumber \\
BR(\tau \rightarrow \mu ee) &=& \left\{|g^{\tau \mu}_3|^2+|g^{\tau \mu}_4|^2+|g^{\tau \mu}_5|^2+|g^{\tau \mu}_6|^2 
   \right\} BR(\tau \to e\nu\nu),\nonumber \\
BR(\tau \rightarrow e\mu\mu) &=& \left\{|g^{\tau e}_3|^2+|g^{\tau e}_4|^2+|g^{\tau e}_5|^2+|g^{\tau e}_6|^2
   \right\} BR(\tau \to e\nu\nu).
\label{bfracs}
\end{eqnarray}
We have used the fact that $BR(\mu \to e\nu\nu)=1$ in writing these expressions.  The $\mu-e$ conversion rate is given by~\cite{Kuno:1999jp}
\begin{equation}
B_{conv} = \frac{2p_e E_e G_F^2 m_{\mu}^3 \alpha^3 Z_{eff}^4 Q_N^2}{\pi^2 Z \Gamma_{capt}} \left[|g_{R}^{\mu e}|^2
  +|g_{L}^{\mu e}|^2\right],
\label{ueconv}
\end{equation}
where $G_F$ is the Fermi constant, $\alpha$ is the QED coupling strength, and the remaining terms are atomic physics 
constants defined in~\cite{Kuno:1999jp}.  Numerical values for titanium, for which the most sensitive limits 
have been obtained~\cite{Wintz:1998rp}, can be found in~\cite{Chang:2005ag}.

We will present a detailed scan of the anarchic RS parameter space in a later section.  For now, to provide some guidance as 
to what scales these rare decays can probe, we perform a few simple estimates.  We set the 5-D fermion 
Yukawas to the values suggested by 5-D Yukawa anarchy, $Y_e=Y_\mu=Y_\tau=2$.  
We also use the intuition described in the previous section to set the mixing matrix entries to the values 
\begin{equation}
U^{L,R}_{11} =1, \,\, U^{L,R}_{12} = \sqrt{\frac{m_e}{m_\mu}},\,\, U^{L,R}_{13} = \sqrt{\frac{m_e}{m_\tau}},
\end{equation}
and similarly for the remaining rows of $U_{L,R}$; for this estimate, we set the phases of these elements to zero.  
We choose a value of $kr_c=11.27$.  We include the first 3 KK modes in this estimate, and we have 
checked that adding more does not affect our results.  Employing these approximations, we check what 
limits can 
be obtained on $M_{KK}$ from each process.  We impose the following bounds: $BR(\mu \rightarrow 3e) < 10^{-12}$, which is the current PDG 
limit~\cite{Eidelman:2004wy}; $B_{conv} < 6.1 \times 10^{-13}$, which is the strongest constraint obtained by the 
experiment SINDRUM II~\cite{Wintz:1998rp}.  For the rare tau decays, we employ the strongest constraints from 
either BABAR or BELLE, which are $BR(\tau \rightarrow l_1\bar{l}_2 l_3) < 2 \times 10^{-7}$ for each mode~\cite{Bfacflavor}.  
We present the bounds on $M_{KK}$ for both the brane Higgs model and the bulk Higgs scenario with $\nu=0$ in Table~\ref{table1}.
\begin{table}[htbp]
\begin{center}
\begin{tabular}{|c|c|c|}
\hline\hline
& Brane Higgs & $\nu=0$ \\ \hline\hline
$BR(\mu \to 3e)$ & 2.5 TeV & 2.0 TeV \\ \hline
$B_{conv}$ & 5.9 & 4.7 \\ \hline
$BR(\tau \to 3\mu)$ & 0.40 & 0.33 \\ \hline
$BR(\tau \to \mu ee)$ & 0.36 & 0.30 \\ \hline
$BR(\tau \to 3e)$ & 0.10 & 0.09 \\ \hline
$BR(\tau \to e\mu\mu)$ & 0.09 & 0.08 \\ \hline\hline
\end{tabular}
\caption{\label{table1} Constraints on the first KK mode mass, $M_{KK}$, coming from various measurements 
for both a brane Higgs field and for the bulk Higgs case with $\nu=0$.  The bounds on $M_{KK}$ are in TeV.}
\end{center}
\end{table}
The limits from $BR(\mu \rightarrow 3e)$ and $B_{conv}$ already probe the multi-TeV region, similar to that possible 
at the LHC.  Although the limits from rare $\tau$-decays are lower, they probe different model parameters which describe 
the third generation.  These bounds will also improve 
as the $B$-factories acquire more data.  We will show that these bounds are generic throughout the entire parameter 
space in a later section.

\section{Dipole operator mediated decays}
\label{dipsection}

\begin{figure}[ht]
\includegraphics[width=0.50\textwidth]{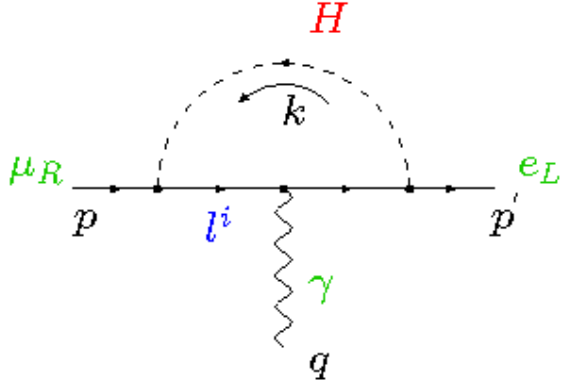}
\caption{The Feynman diagram generating the dipole operator which mediates $l \to l^{'}\gamma$ decays.  $l^i$ are the
physical KK leptons.  We have specialized to $\mu \to e\gamma$ in the figure.  There is a similar diagram with $L\leftrightarrow R$.}
\label{f1}
\end{figure}

We now compute the decays of the form $l \to l^{'}\gamma$, which are induced at the loop level by 
the diagram shown in Fig.~\ref{f1}.  For simplicity, we discuss the decay $\mu \to e\gamma$.  It is simple to translate our 
expressions into results for $\tau$ decays.  The dominant contributions to these amplitudes come from exchange 
of a Higgs boson and KK fermions.  This is because these diagrams contain terms proportional to the fourth power of the 
fermion wave-function ratio $f_{e,\mu}=f^{(1)}_{e,\mu}/f^{(0)}_{e,\mu}$.  For $c=1/2$, this ratio is 
$f_{e,\mu}=2\pi kr_c \approx 70$; it grows rapidly for $c>1/2$, the values relevant for the muon and the electron.  
This strong dependence on the fermion wave-function was first noted in~\cite{Davoudiasl:2000my}.  There are also 
contributions coming from loops of KK $Z$ bosons and KK fermions.  However, as argued in reference~\cite{APS}
for the case of the KK gluon contribution to 
radiative quark decays, the flavor structure
of this diagram is approximately aligned with the $4D$ Yukawa
matrix
and hence gives a suppressed contribution.   
The KK fermion-Higgs diagrams have a different flavor structure than 
the $4D$ Yukawa matrix.

The amplitude for the diagram in Fig.~\ref{f1} is
\begin{eqnarray}
A(\mu\rightarrow e\gamma)&=&\sum_i\int \frac{d^4k}{(2\pi)^4}\overline{u}(p^{'})
(i\Lambda_{e^0i})\frac{i(\slsh{\hat{p}^{'}}+M_{KK}^{(i)})}{\hat{p}^{'2}-M_{KK}^{(i),2}}
(ie\gamma^\mu A_\mu)\frac{i(\slsh{\hat{p}}+M_{KK}^{(i)})}{\hat{p}^2-M_{KK}^{(i),2}}(i\Lambda_{i\mu^0})u(p)
\cdot\frac{i}{k^2-m_H^2} \nln\\
 =& &\!\!\!\overline{u}(p^{'})\left[-eA_\mu\sum_i\Lambda_{e^0i}
\int \frac{d^4k}{(2\pi)^4}
\frac{(\slsh{\hat{p}^{'}}+M_{KK}^{(i)})\gamma^\mu(\slsh{\hat{p}}+M_{KK}^{(i)})}
{(\hat{p}^{'2}-M_{KK}^{(i),2})(\hat{p}^2-M_{KK}^{(i),2})(k^2-m_H^2)}\Lambda_{i\mu^0}\right]u(p),
\end{eqnarray}
where $\hat{p}^{(')}=p^{(')}+k$ and $\Lambda_{ij}$ are the Yukawa 
matrices.  We will assume the external lines are massless, which is valid up to 
subleading corrections in $1/f_{e,\mu}$.  We have denoted the KK fermion masses by $M_{KK}^{(i)}$.  At each 
KK level, there are two vector-like fermion pairs for each flavor with masses $M_{KK}^{(1)}$ and $M_{KK}^{(2)}$, as is clear 
from Eq.~\ref{KKfstates}.  The splitting of these masses through mixing 
will be important in evaluating this contribution.  It is straightforward to evaluate 
this integral to find 
\begin{equation}
A(\mu\rightarrow
e\gamma;q^2)=\frac{1}{2m_\mu}\overline{u}(p^{'})\sigma^{\mu\nu}F_{\mu\nu}u(p)\times
(-i)C(q^2),
\end{equation}
where
\begin{equation}
-iC(q^2)=\frac{iem_\mu}{16\pi^2}\sum_i\Lambda_{e^0i}\left\{\int_0^1 dz
\int_0^{1-z}dy\frac{M_{KK}^{(i)}(1-z)}{q^2y(1-y-z)-(1-z)M_{KK}^{(i),2}-zm_H^2}\right\}\Lambda_{i\mu^0}.
\end{equation}
We now set $q^2=0$ to derive
\begin{equation}
C(q^2=0)=\frac{em_\mu}{32\pi^2}\sum_i\Lambda_{e^0i}\frac{I(m_H^2/M_{KK}^{(i),2})}{M_{KK}^{(i)}}\Lambda_{i\mu^0},
\label{muegamp}
\end{equation}
where $I(x)=1-x+\mathcal{O}(x^2)$.  The branching fraction becomes~\cite{Chang:2005ag}
\be
B(\mu\rightarrow e\gamma)=\frac{12\pi^2}{(G_Fm_\mu^2)^2}
\left[|C_L(0)|^2+|C_R(0)|^2\right],
\label{muegbr}
\ee
where we have inserted the helicity labels $L,R$ on $C$.  These helicity labels dictate which 
elements of the Yukawa matrix $\Lambda$ should be used; we will make this explicit in the following discussion.  
We now consider separately the brane and bulk Higgs field cases.  We will find that the brane Higgs prediction 
for $l \to l^{'}\gamma$ is not calculable because it is sensitive to cut-off scale physics, while for 
the bulk Higgs case we can use our $5D$ effective field theory to make robust predictions.

\subsection{UV sensitivity for the case of brane Higgs field}

The leading contribution in Eq.~\ref{muegamp}, with $m_H=0$ and $I(x)=1$, vanishes up to factors suppressed by $1/f^2$ for a 
brane Higgs field because of the Yukawa matrix structure.  
%
%
With $m_H =0$, we are only considering
contributions proportional to $1 / M^{(1,2)}_{KK}$.  This mass splitting is cancelled by shifts in the 
Yukawa couplings to all orders in $v/M_{KK}$.  The leading result therefore comes 
from $1 / (M^{(1,2)}_{KK})^3$ contributions, and we must consider the $m_H^2$ terms to obtain these.  
The diagonalization of the fermion mass matrix 
in Eq.~\ref{massbulk} is discussed 
in the detail in the Appendix.  The result of this analysis is the following mass splitting:
\begin{equation}
\frac{1}{(M^{(1)}_{KK})^3}-\frac{1}{(M^{(2)}_{KK})^3}=-\frac{3\Delta_1}{M_{KK}^4}.
\end{equation}
This yields the following coefficients of the dipole operator:
\bea
C_L(0)&=&\frac{em_\mu m_H^2}{32\pi^2M_{KK}^4} \left[\Delta_R\Delta_1\Delta_L\right]_{e\mu}, \nonumber \\
C_R(0)&=&\frac{em_\mu m_H^2}{32\pi^2M_{KK}^4} \left[\Delta_R\Delta_1\Delta_L\right]^\dagger_{e\mu}.
\eea
The Yukawa structures entering $C_L$ and $C_R$ differ by a hermitian conjugate.

However, it turns out that this
result is masked by cut-off effects.  A similar ultraviolet sensitivity of Higgs-fermion KK loops  
was also noted in~\cite{APS}.  The expected one-loop contribution from a given 
set of KK modes is finite with size
\begin{equation}
\frac{C^{ KK }_{ L, R } }{ m_{ \mu }^2 } \sim  
\frac{ \lambda_{ 5D }^2 } 
{ 16 \pi^2 } \frac{1}{ M_{ KK }^2 }.
\label{kkest}
\end{equation}
For simplicity, we have not included the relevant mixing matrix elements in this estimate.  Although 
the actual one-loop result for a brane Higgs field vanishes for $m_H=0$, we cannot find a symmetry that 
requires this, and we expect it to be an accident of the one-loop result.  The sum over two independent KK modes would have given a 
logarithmic divergence at one-loop: 
\begin{equation}
\frac{ C^{ KK }_{ L, R } }{ m_{ \mu }^2 } \propto
\log N_{ KK } \sim \log \left( \Lambda_{ 5D } / k \right) \sim
\log \left( \tilde{ \Lambda }_{ 5D } / M_{ KK }\right).
\end{equation}
Here, $N_{ KK } $ is the total number of KK modes in the $5D$ effective theory, $\Lambda_{ 5D }$ is the $5D$ cut-off
of order $10^{19}$ GeV, and $\tilde{ \Lambda }_{ 5D }$ is the warped-down
$5D$ cut-off of order TeV.  Similarly, $M_{ KK }$ is roughly the warped-down curvature scale $k$.  
To obtain this logarithmic divergence, it is crucial that KK fermion-Higgs couplings 
in the sum are independent of the KK index.  We expect that higher-loop contributions
are strongly power divergent because of the increasing number of sums over KK modes, and 
are as important as the one-loop result provided the cut-off scale physics is strongly coupled.

This divergence structure can be more easily seen using power-counting in the $5D$ theory.  Since the 
$5D$ Yukawa coupling has mass dimension $[\lambda_{ 5D } / k]=[-1]$, the loop expansion 
for $\mu \to e\gamma$ has the form
\begin{eqnarray}
\frac{ C^{ KK }_{ L, R } }{ m_{ \mu }^2 } & \sim & \frac{1}{ 16 \pi^2 }
\left( \frac{ \lambda_{ 5D } }{ M_{ KK } } \right)^2 \Big[ \log 
\left( \frac{ \Lambda_{ 5D } }{k} \right) + \frac{1}{ 16 \pi^2 }
\frac{ \lambda_{ 5D }^2 }{ k^2 } \Lambda_{ 5D }^2 +
... \Big].
\end{eqnarray}
In this expression, we have replaced the scale $k \sim 10^{18}$ GeV 
by its warped-down value $M_{ KK }$ in the overall coefficient.  
By simple dimensional analysis, the one-loop
contribution can be log-divergent and the two-loop contribution
is quadratically divergent; in KK language, the power divergence at 
two loops can be seen from the independent sums over $4$ KK modes.  The two-loop result is comparable to the 
one-loop prediction if the cut-off physics is
strongly-coupled: $
\Lambda_{ 5D }^2 / k^2 \times 
\lambda_{ 5D }^2 / \left( 16 \pi^2 \right) \sim 1$.  Therefore, the KK loop contribution is not calculable in this case.

Based on the above discussion, we also expect the higher-dimensional
operators in the $5D$ theory coming from physics at the cut-off scale to
be important.  The relation between the warped-down $5D$ cut-off in the Yukawa sector and the KK scale for a brane Higgs field is 
$\tilde{ \Lambda }_{ 5D } \sim M_{ KK } / \left( 4 \pi / \lambda_{ 5D } \right),$  based on power counting of the $5D$ loop factor.
To obtain the cut-off operator,
we replace $M_{ KK }$ in Eq.~\ref{kkest}
by the cut-off scale $\tilde{ \Lambda }_{ 5D }$, 
and the loop factor by $\sim 1$, since the cut-off
effect has no loop suppression.  This shows that
$\mu \rightarrow e \gamma $ is an UV sensitive observable
for a Higgs field on the TeV brane. We can only parameterize the contribution as: 
\begin{equation}
C^{ total }_{L , R} =  a \frac{ m_{ \mu }^2 }{ \tilde{ \Lambda }_{ 5D }^2 } 
\times U^{L, R}_{ 12 },
\end{equation}
where $a$ is an unknown, $O(1)$ coefficient, and we have included the appropriate mixing matrix element.

We now show that we can reliably calculate dipole induced decays for a bulk Higgs field.  
The Yukawa coupling in this case has mass dimension $[\lambda_{ 5D } / \sqrt{k}]=[-1/2]$, so the 
loop expansion is instead
\begin{eqnarray}
\frac{ C^{ KK }_{ L , R }}{ m_{ \mu }^2 }& \sim & \frac{1}{ 16 \pi^2 } 
\left( \frac{ \lambda_{ 5D } }{ \sqrt{ M_{ KK } } } \right)^2 
\Big[ \frac{1}{ M_{ KK } } + \frac{1}{ 16 \pi^2 }
\frac{ \lambda_{ 5D }^2 }{ M_{ KK } } \log \left(
\frac{ \Lambda_{ 5D } }{k} \right) + \left( \frac{1}{ 16 \pi^2 } 
\right)^2 \frac{ \lambda_{ 5D }^4 }{ M_{ KK }^2 } 
\tilde{ \Lambda }_{ 5D }  +...\Big]. \nonumber
\end{eqnarray}
From this $5D$ power-counting, we see 
that the one-loop KK contribution is finite.  The two-loop result 
is logarithmically divergent, but is smaller than the 
one-loop prediction by $\sim 0.1$ provided $\lambda_{ 5D } \stackrel{<}{\sim} 4$.  
Three-loop and higher contributions are power-divergent
and comparable to the two-loop result, but are again smaller than the
one-loop effect.

Thus, in the bulk Higgs case,
the KK effect is calculable.  The effects from cut-off scale operators are suppressed, and 
we can reliably make a prediction using the RS theory.  In our numerical analysis, we will 
include dipole decays for the bulk Higgs field case.  For the brane Higgs scenario we will 
simply neglect them, since we cannot make a reliable prediction.

\subsection{Contributions from a bulk Higgs field}

We now consider the scenario when the Higgs boson is allowed to propagate in the bulk.  In this case, 
the KK mode result is not overwhelmed by cut-off scale operators.  The $m_H=0$ limit does 
not vanish for a bulk Higgs.  We make this approximation in our discussion, since the corrections 
are $\mathcal{O}(m_H^2/M_{KK}^2)$.  We first work out the Yukawa structure appearing in Eq.~\ref{muegamp}.  
Using the results in the Appendix for the two KK fermions appearing in the diagram of Fig.~\ref{f1}, we find
\begin{eqnarray}
(\overline{e}^0_Ll_R^1)(\overline{l}^1_L\mu_R^0) &=&
\left[\Delta_R\left[1+\left(\frac{X}{4}-\frac{\Delta_2}{M_{KK}}\right)\right]\right]_{el}
\left(\frac{1}{M^{(1)}_{KK}}\right)
\left[\left[1+\left(\frac{X}{4}-\frac{\Delta_2}{M_{KK}}\right)\right]\Delta_L\right]_{l\mu} \nonumber \\
(\overline{e}^0_Ll_R^2)(\overline{l}^2_L\mu_R^0) &=&
\left[\Delta_R\left[1-\left(\frac{X}{4}-\frac{\Delta_2}{M_{KK}}\right)\right]\right]_{el}
\left(-\frac{1}{M^{(2)}_{KK}}\right)
\left[\left[1-\left(\frac{X}{4}-\frac{\Delta_2}{M_{KK}}\right)\right]\Delta_L\right]_{l\mu}.
\end{eqnarray}
In this expression we must sum over $l=e,\mu,\tau$.  To simplify this we use the splitting between the 
KK fermion masses derived in the Appendix:
\begin{equation}
\frac{1}{M^{(1)}_{KK}}-\frac{1}{M^{(2)}_{KK}}=-\frac{\Delta_1+\Delta_2}{M_{kk}^2} +
\mathcal{O}\left(\frac{v^3}{M_{kk}^4}\right).
\end{equation}
We find the following results for the dipole operator coefficients:
\begin{eqnarray}
C_L(0)&=&\frac{3em_\mu}{32\pi^2M_{KK}^2}\left[\Delta_R\Delta_2\Delta_L\right]_{e\mu} \nonumber \\
C_R(0)&=&\frac{3em_\mu}{32\pi^2M_{KK}^2}\left[\Delta_R\Delta_2\Delta_L\right]^\dagger_{e\mu}
\label{dipcoeffs}
\end{eqnarray}
We note that in the limit of the Higgs boson being localized on the TeV brane, $\Delta_2 \to 0$; the 
result vanishes in this limit, as required.

An identical analysis can be performed for $\tau\rightarrow\mu\gamma$ and
$\tau\rightarrow e\gamma$.  We simply replace $m_{\mu} \to m_{\tau}$, change 
the indices of the Yukawa structure appropriately in Eq.~\ref{dipcoeffs}, and normalize the expression 
to the decay $\tau \to e\nu\nu$.  We now 
perform an estimate of the bounds similar to that performed in the brane Higgs case.  We set 
$Y_e=Y_{\mu}=Y_{\tau}=2$, and set the mixing matrix elements to their canonical values as described before.  
We also set $\nu=0$.  We impose the following bounds on each of the three dipole decays:
$BR(\mu \to e\gamma) < 1.2 \times 10^{-11}$, as obtained from~\cite{Eidelman:2004wy}; 
$BR(\tau \to \mu\gamma) < 9 \times 10^{-8}$, the stronger of the bounds coming from BABAR and BELLE~\cite{taumug};
$BR(\tau \to e\gamma) < 1.1 \times 10^{-7}$, again the stronger of the bounds coming from BABAR and BELLE~\cite{taueg}.
We find the following constraints for the canonical parameters:
\begin{eqnarray}
BR(\mu \rightarrow e\gamma)&:& M_{KK} > 15.8 \, {\rm TeV}; \nonumber \\
BR(\tau \rightarrow e\gamma)&:& M_{KK} > 0.9 \, {\rm TeV}; \nonumber \\
BR(\tau \rightarrow \mu\gamma)&:& M_{KK} > 1.6 \, {\rm TeV}.
\label{gammatestbounds}
\end{eqnarray}
The constraints, particularly from $BR(\mu \to e\gamma)$, are quite strong.  This arises in part from the large 
value of the Yukawa coupling, $Y=2$, as we now discuss.

\subsection{Tension between tree-level and loop-induced processes}

We now discuss a tension between processes caused by tree-level gauge boson mixing such as $\mu-e$ conversion 
and $l \to l_1 \bar{l}_2 l_3$, and dipole operator decays.  These have opposite dependences on the $5D$ Yukawa 
couplings, leading to strong constraints for all parameter choices.  We first give a very simple scaling argument 
to motivate this, and then present numerical proof.

\begin{figure}[htbp]
\centerline{
\psfig{figure=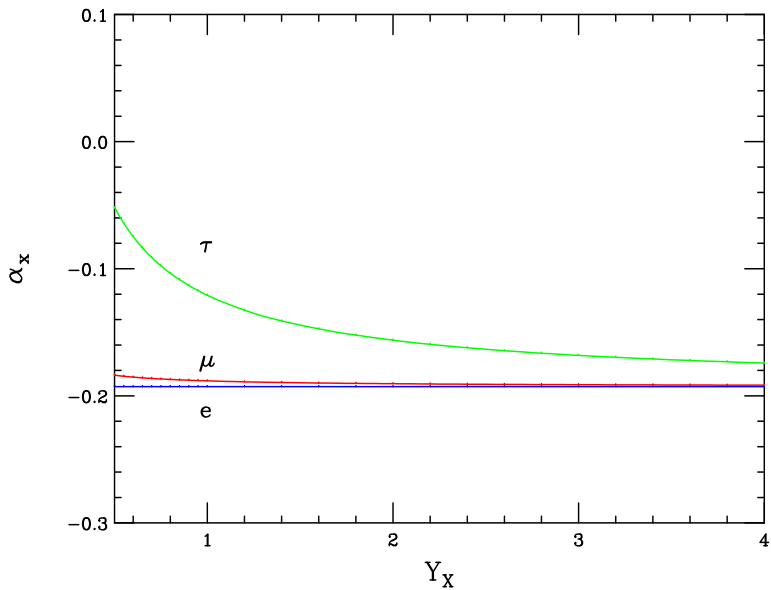,height=10.2cm,width=7.8cm,angle=90}}
\caption{The ratios of the zero-mode fermion couplings to $Z^{(1)}$ over their SM values, for $x=e,\mu,\tau$, as functions of the
Yukawa couplings $Y_x$.}
\label{alphaplot}
\end{figure}

Our scaling argument uses the dependence of each process on the zero-mode fermion wave-function $f_{l}^{(0)}$ 
evaluated at the TeV brane.  We will work for simplicity 
in the large $\nu$ limit, which mimics a brane-localized Higgs field.  From Eqs.~\ref{zerowf} and~\ref{bulklambda}, we find that the 
wave-function scales roughly as $f_{l}^{(0)} \sim 1/\sqrt{\lambda_{5D}}$.  The wave-function has weak $c$-dependent factors which we will 
ignore in this argument.  The quantity that governs the flavor violation in gauge boson mixing is 
the difference between $\alpha_l$'s, as is clear from 
Eq.~\ref{eqexam}.  In the definition of $\alpha_l$ in Eq.~\ref{alphadef}, 
we can divide the overlap integral into two regions, one near the Planck brane and the other 
near the TeV brane, to show that the former is $c$-independent
and that the latter carries the $c$-dependence and must be
$\alpha_l |_{ non-universal } \sim [f_{l}^{(0)}] |_{TeV brane}^2 \sim 1/\lambda_{5D}$. 
We therefore expect the non-universal part of $\alpha_l$, and hence the flavor violation, 
to decrease for larger Yukawa couplings, which is indeed what we observe in Fig.~\ref{alphaplot}.  For 
the dipole mediated decays, recall that in Section~\ref{dipsection} we claimed that the operator coefficients $C_{L,R}$ scaled 
as $C_{L,R} \sim 1/[f_{l}^{(0)}]^4 \sim \lambda_{5D}^2$; this can be verified using Eq.~\ref{dipcoeffs} and the results in the Appendix.  
The constraints coming from $l \to l^{'}\gamma$ decays will increase with larger Yukawa couplings, the opposite dependence of the 
tree-level processes.

%

To exhibit this behavior we present in Table~\ref{table2} the bounds on the first KK mode mass for canonical mixing 
angles, $\nu=0$, and for the two choices of Yukawa strength $Y_e=Y_{\mu}=Y_{\tau}=1,2$.  We show the two most constraining 
processes, $\mu-e$ conversion and $BR(\mu \to e\gamma)$.  The dependence on the Yukawa couplings agrees with our simple estimate 
above.  We will find in the next section that this leads to strong constraint throughout the entire model parameter space.

\begin{table}[htbp]
\begin{center}
\begin{tabular}{|c|c|c|}
\hline\hline
& $Y=1$ & $Y=2$ \\ \hline\hline
$B_{conv}$ & 6.7 TeV & 4.7 TeV \\ \hline
$BR(\mu \to e\gamma)$ & 8.0 & 15.8 \\ \hline
\end{tabular}
\caption{\label{table2} Constraints on the first KK mode mass, $M_{KK}$, coming from $\mu-e$ conversion 
and $BR(\mu \to e\gamma)$, for canonical mixing angles, $\nu=0$, and for $Y=1,2$.}
\end{center}
\end{table} 

\section{Monte-Carlo scan of the anarchic RS parameter space}

In this section we present our Monte-Carlo scan of the RS parameter space, to determine in detail how well the RS geometric origin of 
flavor can be tested by current and future lepton flavor-violation experiments.  

We first describe the ranges over which we scan the various RS parameters.  The scenario introduced in the previous sections contains 
the following free parameters: $Y_e$, $Y_\mu$, $Y_\tau$, the overall Yukawa
couplings for the electron, muon, and tau; $U^{L,R}_{ij}$, the 
elements of both the left and right-handed mixing matrices; the KK mass $M_{KK}$.  We make the following assumptions in our scan.
\begin{itemize}
  \item We restrict the Yukawa couplings to the range $Y_x \in [\frac{1}{2},4]$.  As discussed before, the natural value is $Y_x \approx 2$.  
    Values larger than 4 begin to invalidate the perturbative 
    expansion, while values smaller than $1/2$ introduce an unnatural hierarchy in the model.  We explained in the previous section that 
    flavor violation cannot be removed by making the Yukawa couplings either large or small, due to tension between tree-level and 
    loop-induced processes.
  \item We implement the anarchy of 5-D couplings in our scan, which indicates that $U^{L,R}_{ii} \sim 1$, $U^{L,R}_{12} \sim \sqrt{m_e/m_\mu}$, 
    $U^{L,R}_{13} \sim \sqrt{m_e/m_\tau}$, etc.  We fix $U^{L,R}_{ii}=1$, and define the canonical values
    \begin{equation}
      U^{c}_{12} = \sqrt{\frac{m_e}{m_\mu}},\,\, U^{c}_{13} = \sqrt{\frac{m_e}{m_\tau}},\,\, U^{c}_{23} = \sqrt{\frac{m_\mu}{m_\tau}}.
    \end{equation}
    We then vary $U^{L}_{12}=\beta^{L}_{12} U^{c}_{12}$, with $\beta^{L}_{12} \in [1/4,4]$.  We independently 
    vary $U^{R}_{12}$, $U^{L,R}_{13}$, and $U^{L,R}_{23}$ in a similar fashion.  Again, we restrict the values to these ranges to insure 
    no unnatural hierarchies in model parameters.  We generate phases for the six independent $U^{L,R}$ in the range $[0,2\pi]$.
  \item We approximately implement unitarity of the mixing matrices by setting $U^{L,R}_{21} = -\left(U^{L,R}_{12}\right)^{*}$, etc.  This assures that
    unitarity is maintained up to corrections of the level $\sqrt{m_e/m_\mu}$, $\sqrt{m_\mu/m_\tau}$, which is 
    sufficient for the scan performed here.
\end{itemize}
We scan over the following fifteen independent parameters: the three $Y_x$, and the six complex mixing matrix elements $U^{L,R}_{12}$, 
$U^{L,R}_{13}$, and $U^{L,R}_{23}$.  We generate 1000 sets of fifteen random
numbers, and distribute them in the ranges indicated above for fixed $M_{KK}$.  We perform two separate scans, one for a brane Higgs 
field and one for a bulk Higgs with $\nu=0$.  The $\nu$ dependence of the bulk Higgs field bounds is studied separately. 

\subsection{Scan for the brane Higgs field scenario}

We first perform a Monte-Carlo scan of the parameter space of the brane 
Higgs scenario.  As discussed in Section~\ref{dipsection}, 
dipole decays of the form $l \to l^{'}\gamma$ are UV sensitive.  
We do not consider these decays in the brane Higgs case, which leaves us 
with $\mu-e$ conversion, $\mu \to 3e$, and $\tau \to l_1 \bar{l}_2 l_3$.

\begin{figure}[htb]
\centerline{
\psfig{figure=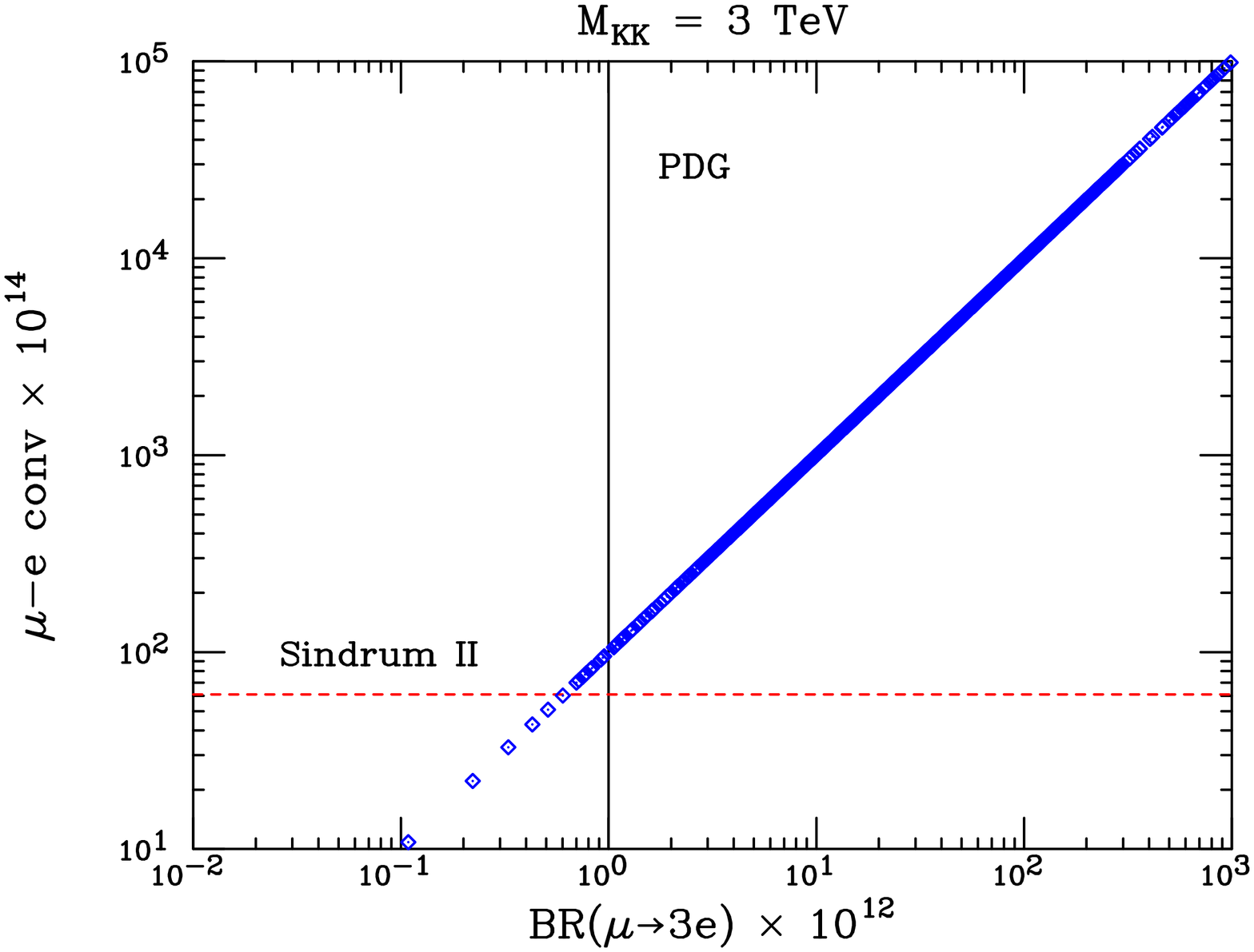,height=8.0cm,width=6.1cm,angle=90} \hspace{0.15cm} 
\psfig{figure=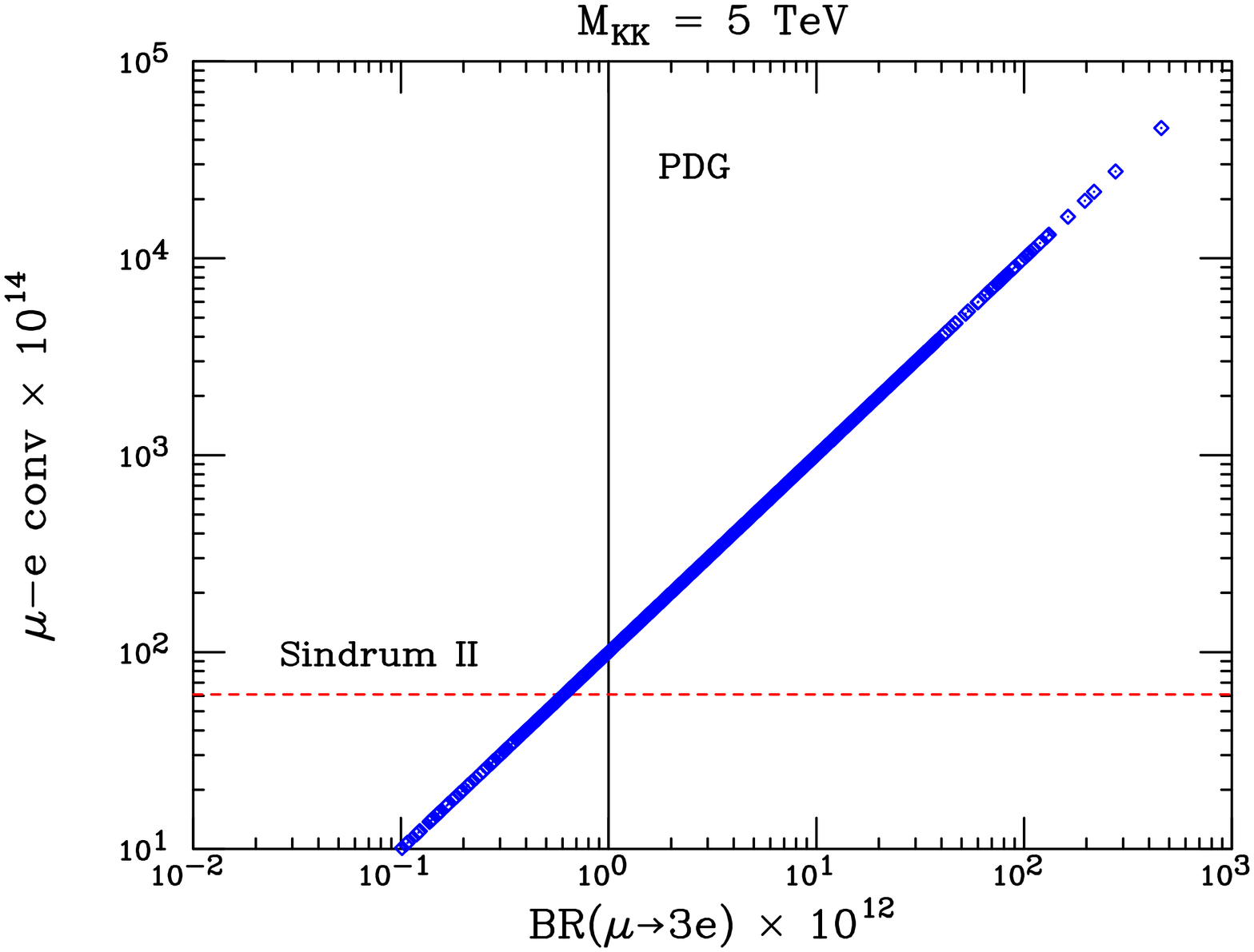,height=8.0cm,width=6.1cm,angle=90}}
\vspace{0.05cm}
\centerline{
\psfig{figure=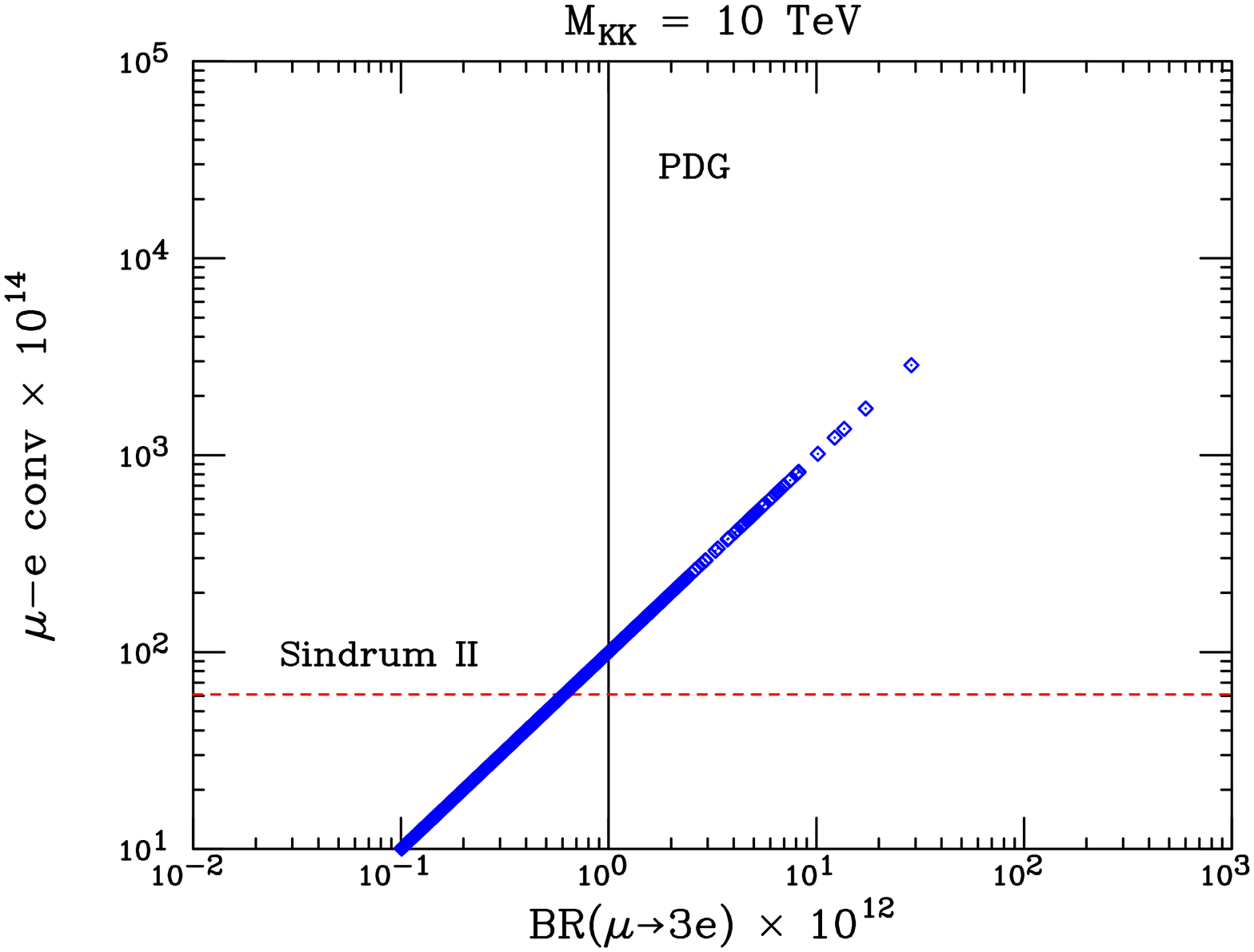,height=8.0cm,width=6.1cm,angle=90}}
\vspace{-0.2cm}
\caption{Scan of the $\mu \rightarrow 3e$ and $\mu-e$ conversion predictions for $M_{KK}=3,5,10$ TeV.  The solid and dashed lines 
are the PDG and SINDRUM II limits, respectively.}
\label{brane:mubounds}
\end{figure}

We first study the muonic processes $\mu \rightarrow 3e$ and $\mu-e$ conversion.  We show in Fig.~\ref{brane:mubounds} scatter plots of 
the predictions for $BR(\mu \rightarrow 3e)$ and $B_{conv}$ coming from our scan of the RS parameter space, for the $KK$ scales 
$M_{KK}=3,5,10$ TeV.  The most sensitive probe is the SINDRUM II limit of $B_{conv}<6.1 \times 10^{-13}$~\cite{Wintz:1998rp}.  This rules 
out a large fraction of the parameter space for $M_{KK}<5$ TeV, and restricts the allowed parameters even at 10 TeV.  The PDG 
limit of $BR(\mu \rightarrow 3e) < 10^{-12}$ is less severe: although it rules out a large fraction of the $M_{KK}=3$ TeV parameter space, most 
of the $M_{KK}=5$ TeV space is still allowed.  We note there is an almost perfect correlation between the RS predictions
for the two processes.  This is not surprising; it is clear from Eqs.~\ref{bfracs} and~\ref{ueconv} that they depend almost 
identically on the same mixing angles.

This result has implications for both the aesthetic appeal of the anarchic RS flavor picture, and the observation of this physics
at the LHC.  Although points with $M_{KK} \leq 3$ TeV are still allowed, it is clear from Fig.~\ref{brane:mubounds} 
that the model as formulated in our scan prefers KK masses of 5 TeV or larger.  Increasing the KK scale to these 
higher values introduces a large fine-tuning in the electroweak symmetry breaking sector 
and is therefore not favored~\cite{ADMS,Agashe:2004rs}.  With such large KK masses, many associated states will also be too heavy to 
observe at the LHC.  The other method of avoiding these constraints, reducing the $U^{L,R}_{ij}$ matrix elements to
the appropriate level, implies either some additional structure or fine-tuning in the 5-D Yukawa matrix.  We have studied 
the minimal model here, and it seems likely that more structure in the 5-D Yukawa matrix is needed for a completely natural description 
of the first and second generation flavor pattern in the brane Higgs case.

\begin{figure}[htb]
\centerline{
\psfig{figure=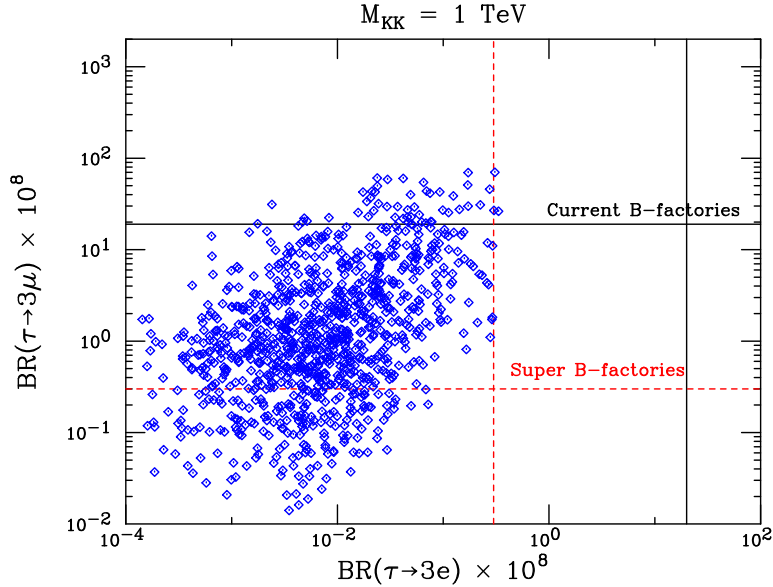,height=10.2cm,width=7.8cm,angle=90}}
\vspace{-0.2cm}
\caption{Scan of the $\tau \rightarrow 3e$ and $\tau \rightarrow 3\mu$ predictions for $M_{KK}=1$ TeV.  The solid and dashed lines 
are the current $B$-factory and projected super-$B$ factory limits, respectively.}
\label{brane:tauplot}
\end{figure}

Another sector of the RS flavor picture to explore is that involving the third generation $\tau$.  This tests different model 
parameters than the muonic processes.  We show in Fig.~\ref{brane:tauplot} a scatter plot of the RS predictions
for $BR(\tau \rightarrow 3e)$ and $BR(\tau \rightarrow 3\mu)$ for $M_{KK}=1$ TeV, together with 
the best limits coming from BABAR and BELLE.  The lowest $KK$-scale allowed by electroweak 
precision tests in anarchic RS models is typically a few TeV.  The $B$-factories are beginning to probe this region in 
the mode $\tau \to 3\mu$.  There are plans to build a super-$B$ factory with an integrated luminosity approaching 
10 ${\rm ab}^{-1}$~\cite{Hewett:2004tv}.  The projected limits from this experiment are included in Fig.~\ref{brane:tauplot}.  
Both the $\tau \to 3\mu$ and $\tau \to 3e$ modes at a super-$B$ factory will constrain the anarchic RS parameter space.  
The LHC also has sensitivity to rare $\tau$ decays~\cite{Unel:2005fj}; however, the projected sensitivities are slightly 
weaker than the current $B$-factory constraints, and have not been included.  The expected sensitivities to rare $\tau$ 
decays at a future linear collider are also weaker than the limits set by the $B$-factories.  
Although the $M_{KK} \sim 1$ TeV scales 
probed with $\tau \to l_1 \bar{l}_2 l_3$ decays are lower than those constrained by $\mu-e$ conversion and $\mu \to 3e$, we 
stress that different model parameters are tested by each set of processes.

%

\subsection{Scan for the bulk Higgs field scenario}

We now present the results of our scan over the bulk Higgs parameter space.  For the scan we set $\nu=0$, which mimics 
the composite (or $A_5$) Higgs model of~\cite{Agashe:2004rs}; we present separately the $\nu$ dependence of the most important 
constraints. 

\begin{figure}[htb]
\centerline{
\psfig{figure=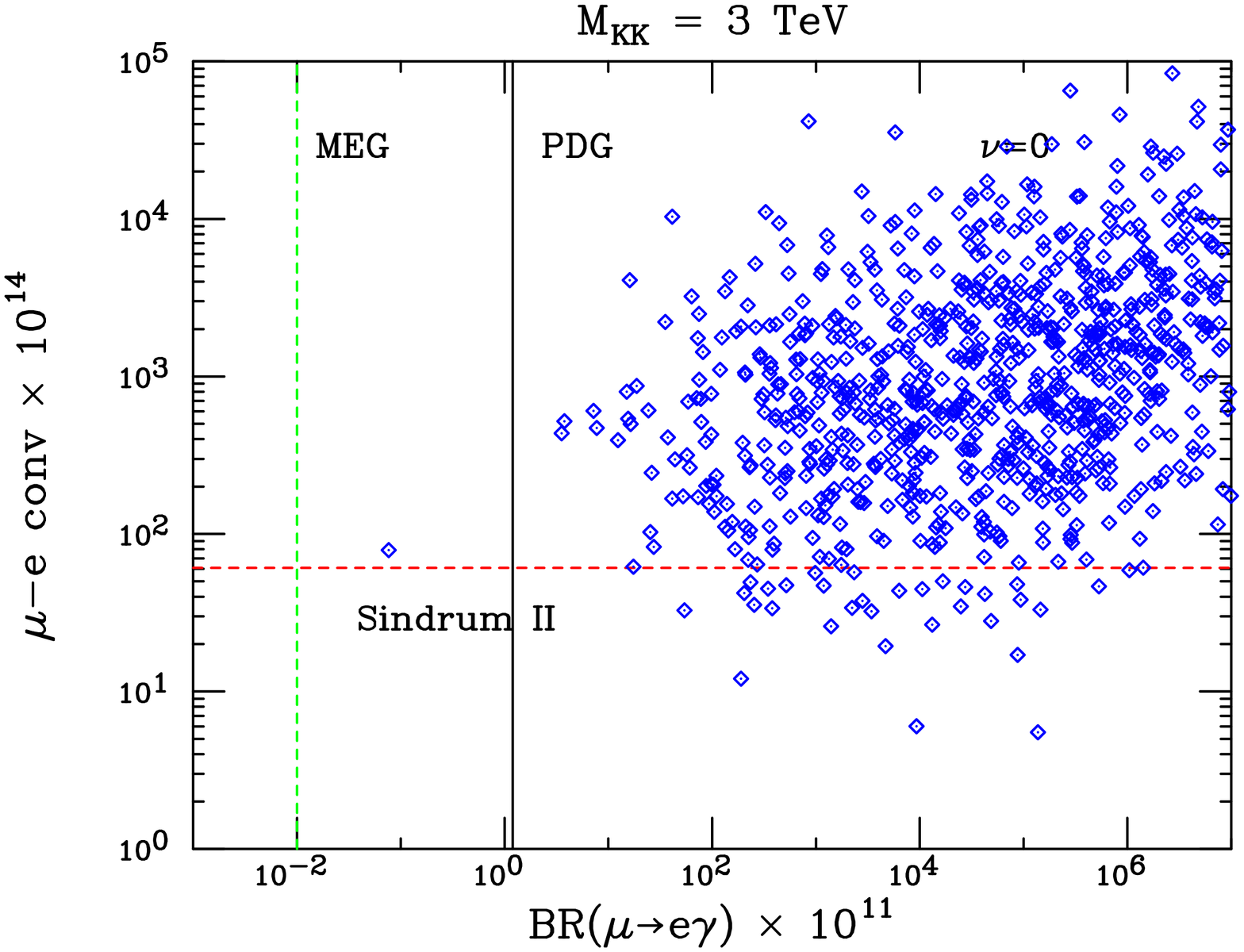,height=8.0cm,width=6.1cm,angle=90} \hspace{0.15cm}
\psfig{figure=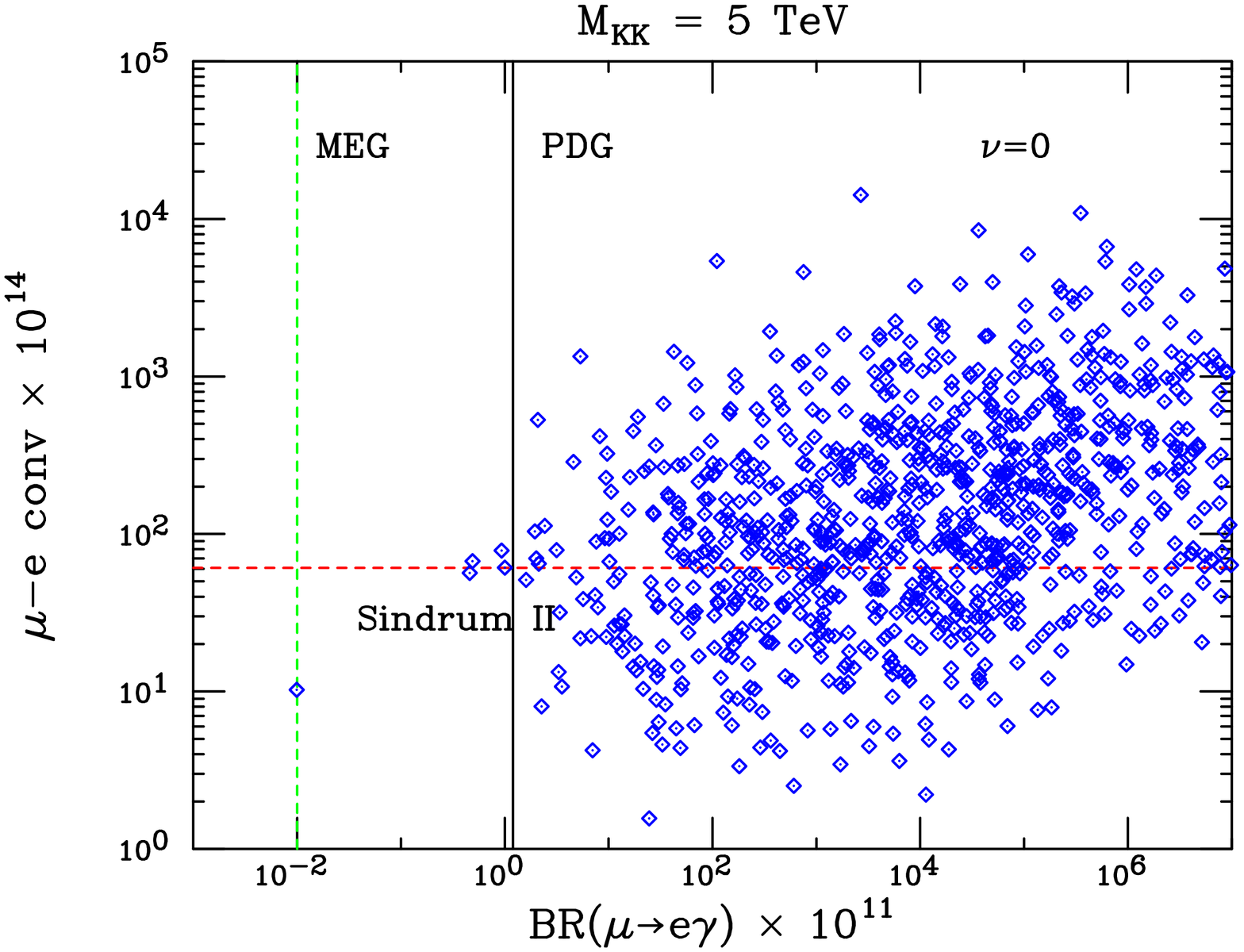,height=8.0cm,width=6.1cm,angle=90}}
\vspace{0.05cm}
\centerline{
\psfig{figure=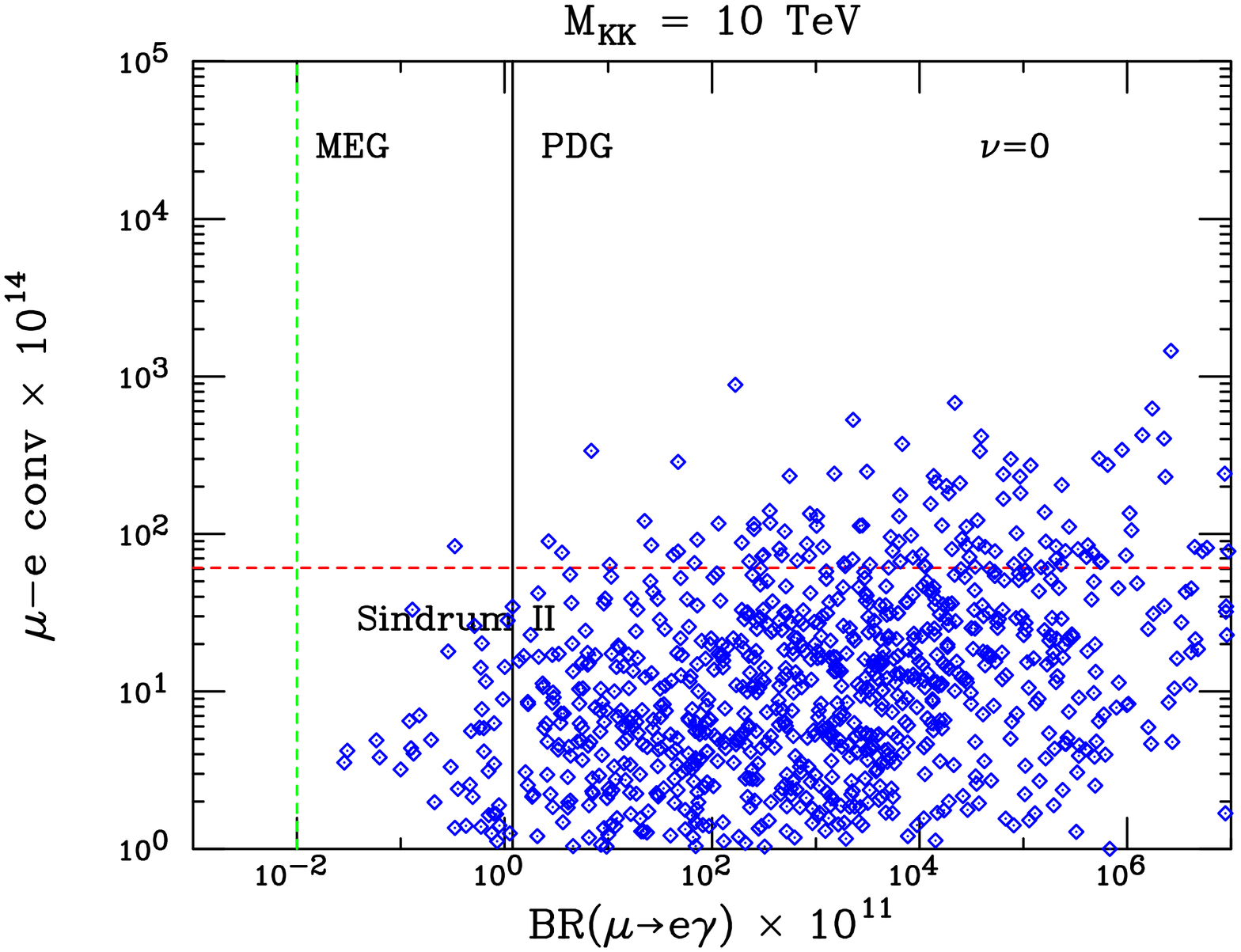,height=8.0cm,width=6.1cm,angle=90}}
\vspace{-0.2cm}
\caption{Scan of the $\mu \rightarrow e\gamma$ and $\mu-e$ conversion predictions for $M_{KK}=3,5,10$ TeV and $\nu=0$.  
The solid line denotes the PDG bound on $BR(\mu \to e\gamma)$, while the dashed lines indicate the SINDRUM II limit on $\mu-e$ 
conversion and the projected MEG sensitivity to $BR(\mu \to e\gamma)$.}
\label{A5:mubounds}
\end{figure}

\begin{figure}[htb]
\centerline{
\psfig{figure=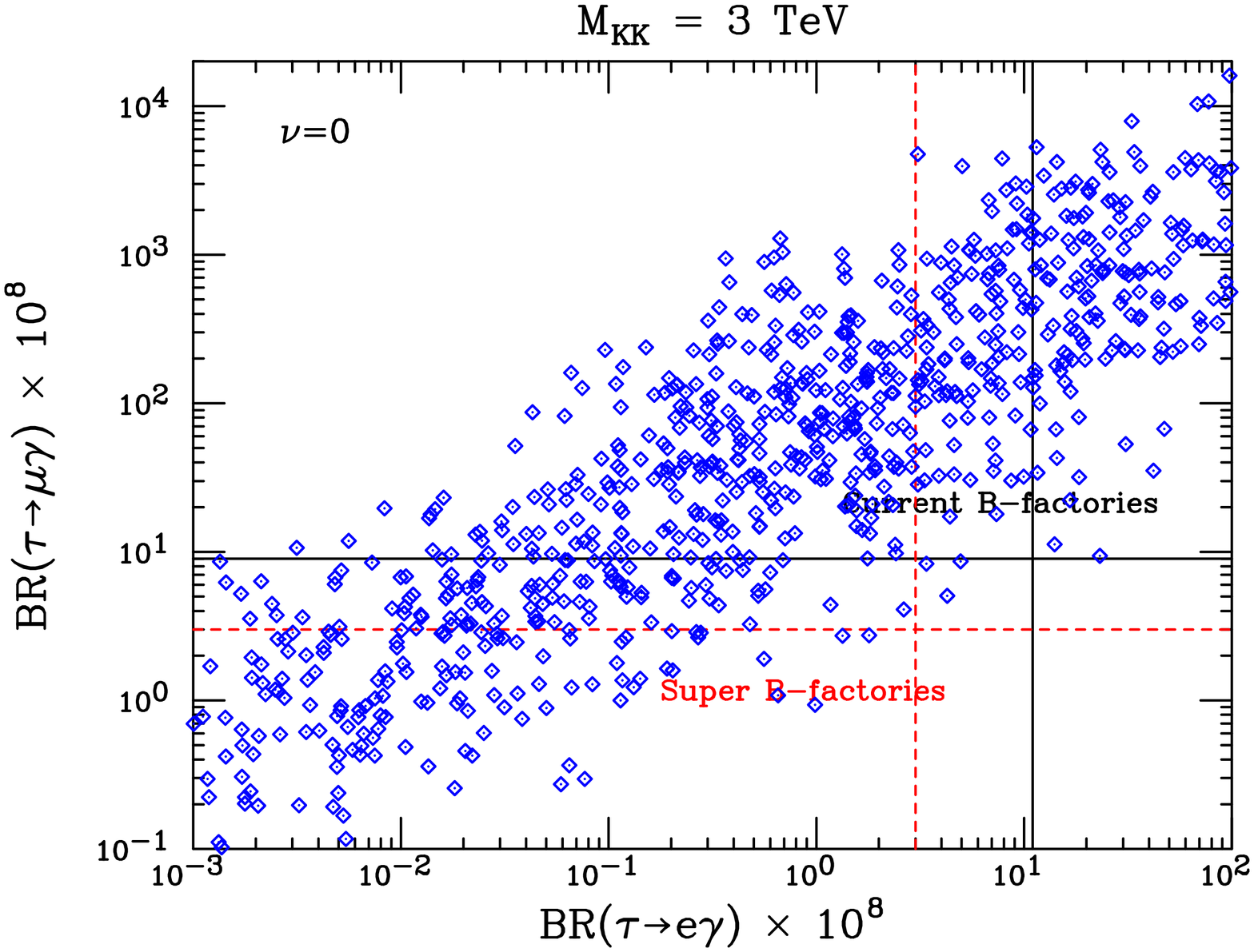,height=10.2cm,width=7.8cm,angle=90}}
\vspace{-0.2cm}
\caption{Scan of the $\tau \rightarrow \mu\gamma$ and $\tau \rightarrow e\gamma$ predictions for $M_{KK}=3$ TeV and $\nu=0$.  
The solid and dashed lines
are the current $B$-factory and projected super-$B$ factory limits, respectively.}
\label{A5:tauplot}
\end{figure}

We again begin by considering muon initiated processes.  The constraints from $\mu \to 3e$ and $\mu-e$ conversion are 
highly correlated, as we saw in the previous subsection.  Since the bounds from $\mu-e$ conversion are stronger, 
we focus on this and $\mu \to e\gamma$.  We show in Fig.~\ref{A5:mubounds} scatter plots of
the predictions for $BR(\mu \to e\gamma)$ and $B_{conv}$ coming from our scan of the RS parameter space, for the $KK$ scales
$M_{KK}=3,5,10$ TeV.  For $\mu \to e\gamma$ we include both the current constraint from the 
Particle Data Group~\cite{Eidelman:2004wy} 
and the projected sensitivity of MEG~\cite{Signorelli:2003vw}.  
The current bounds from $\mu \to e\gamma$ are quite 
strong; from the $M_{KK}=3$ TeV plot in Fig.~\ref{A5:mubounds},
we see that  
only 
one parameter choice satisfies the $BR(\mu \to e\gamma)$ bound.  
This point does not satisfy the $\mu-e$ conversion constraint.  
We can estimate that it would satisfy both bounds for $M_{KK}>3.1$ TeV.  In our scan 
over 1000 sets of model parameters the absolute lowest scale allowed is thus slightly larger than 3 TeV. 
Also, a large portion 
of the parameter set at both 
5 and 10 TeV conflict with these bounds. 
We again find the need for a KK scale of $M_{KK} \geq 5$ TeV or additional structure 
in the mixing between the first and second generations to satisfy the experimental constraints for a significant 
fraction of model parameter space.  In Fig.~\ref{A5:tauplot} we 
present the anarchic RS predictions for $\tau \to \mu\gamma$ and $\tau \to e\gamma$, together with current and future 
$B$-factory constraints, for $M_{KK}=3$ TeV.  The $\tau \to \mu\gamma$ mode is currently probing the few TeV range, while
$\tau \to e\gamma$ will begin to test the anarchic RS scenario during the running of a super-$B$ factory.

%

\begin{figure}[htb]
\centerline{
\psfig{figure=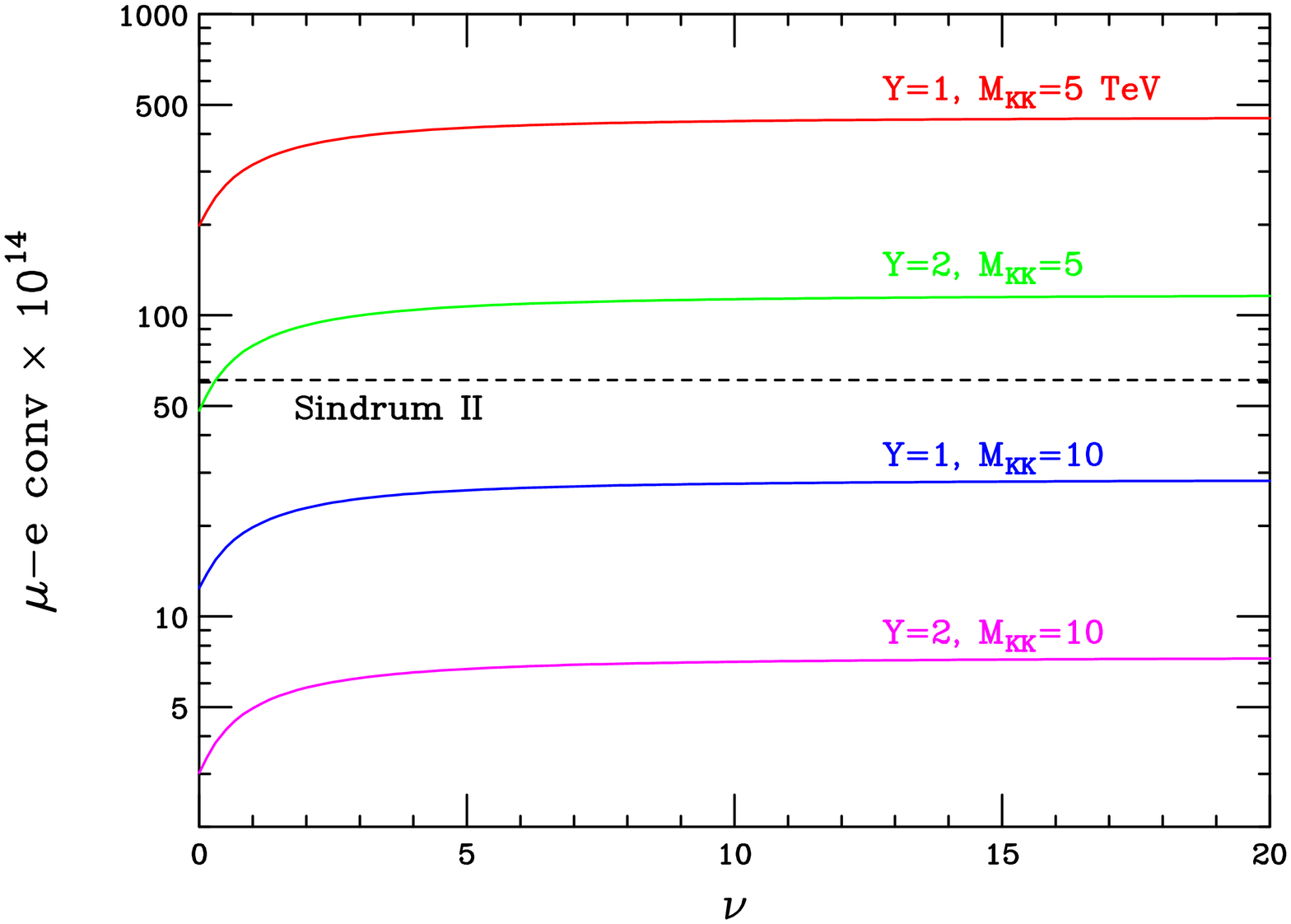,height=8.0cm,width=6.1cm,angle=90} \hspace{0.15cm}
\psfig{figure=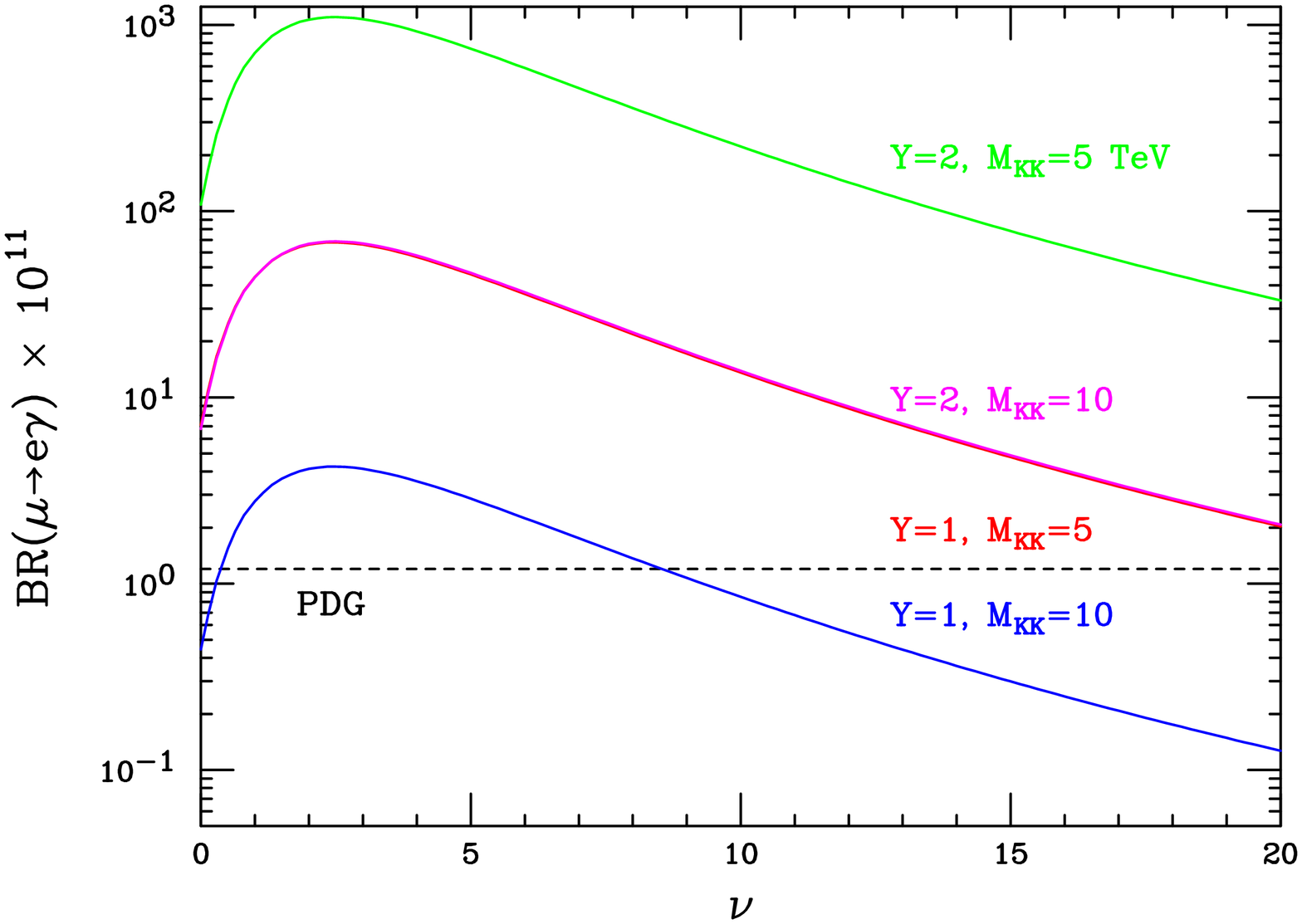,height=8.0cm,width=6.1cm,angle=90}}
\vspace{-0.2cm}
\caption{$\nu$ dependence of the RS predictions for $\mu-e$ conversion and $\mu \to e\gamma$ for 
canonical mixing angles and for several choices of $Y_x$ and $M_{KK}$.  In the right panel, 
the $Y=1$, $M_{KK}=5$ TeV and the $Y=2$, $M_{KK}=10$ TeV lines overlap.}
\label{nudep}
\end{figure}

To study the sensitivity of the bulk Higgs field scenario to the location of the Higgs boson in the fifth dimension, we 
show in Fig.~\ref{nudep} the dependence of the $\mu-e$ conversion rate and $BR(\mu \to e\gamma)$ on $\nu$.  We set 
the mixing angles to their canonical values, and show results for $Y_x=1,2$ and $M_{KK}=5,10$ TeV.  The $\mu-e$ conversion 
results are weakest for $\nu=0$, and quickly asymptote to the brane Higgs result as $\nu$ becomes large.  The variation 
of $\mu \to e\gamma$ with $\nu$ is more intricate.  The vanishing of the calculable component of this process 
as the Higgs boson is moved towards the TeV brane, discussed in Section~\ref{dipsection}, is clearly seen in Fig.~\ref{nudep}.  
However, we expect cut-off effects to become more important for large $\nu$.  There is a strong dependence 
of the process on the position of the Higgs field for small $\nu$, with the result varying by an 
order of magnitude for $0 \leq \nu \leq 5$.  The $\nu=0$ case is again the most favorable choice.  Since UV sensitivity of the 
model is reduced for a bulk Higgs field, and since the experimental constraints are weakest for $\nu=0$, we conclude 
that there is a preference for models of the type presented in~\cite{Agashe:2004rs}.

\subsection{Future sensitivities of MEG and PRIME}

Finally, we emphasize here the importance of future searches for $\mu-e$ conversion by PRIME and 
$\mu \to e\gamma$ by MEG.  Our analysis has shown that with some small tuning of parameters, particularly 
for those describing the mixing of the first and second generation, KK scales of 3 TeV are allowed 
by current measurements.  Alternatively, KK scales of 5 TeV are permitted with completely natural 
parameters.  Super-$B$ factory searches for rare $\tau$ decays will not significantly constrain scales $M_{KK} \geq 5$ TeV.  
The LHC search reach for the new states predicted by the anarchic RS scenario is expected to be 
around 5-6 TeV.  It is therefore difficult to definitively test the RS geometric origin of flavor using 
data from $B$-factories and the LHC.  

\begin{figure}[htb]
\centerline{
\psfig{figure=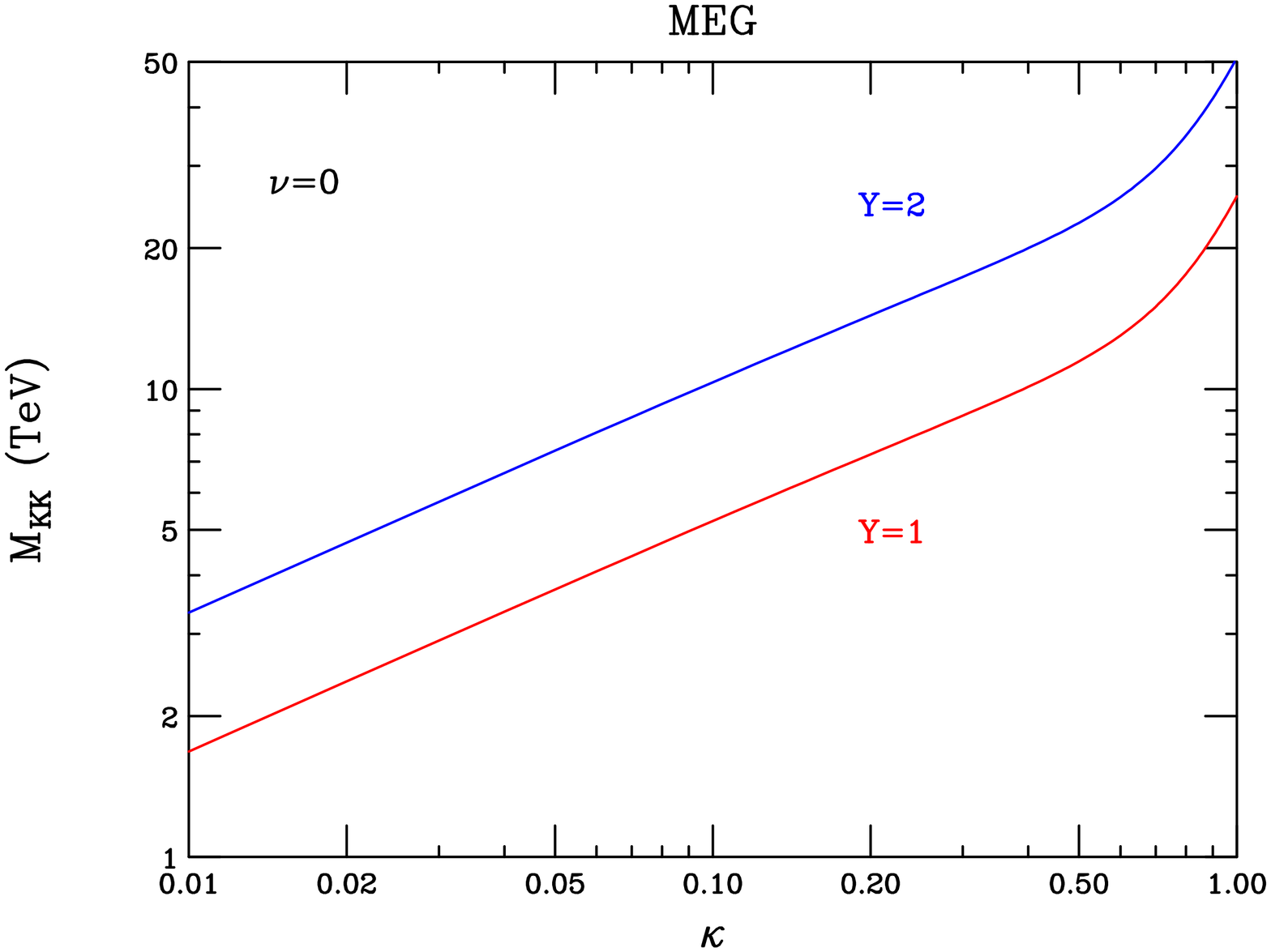,height=8.0cm,width=6.1cm,angle=90} \hspace{0.15cm}
\psfig{figure=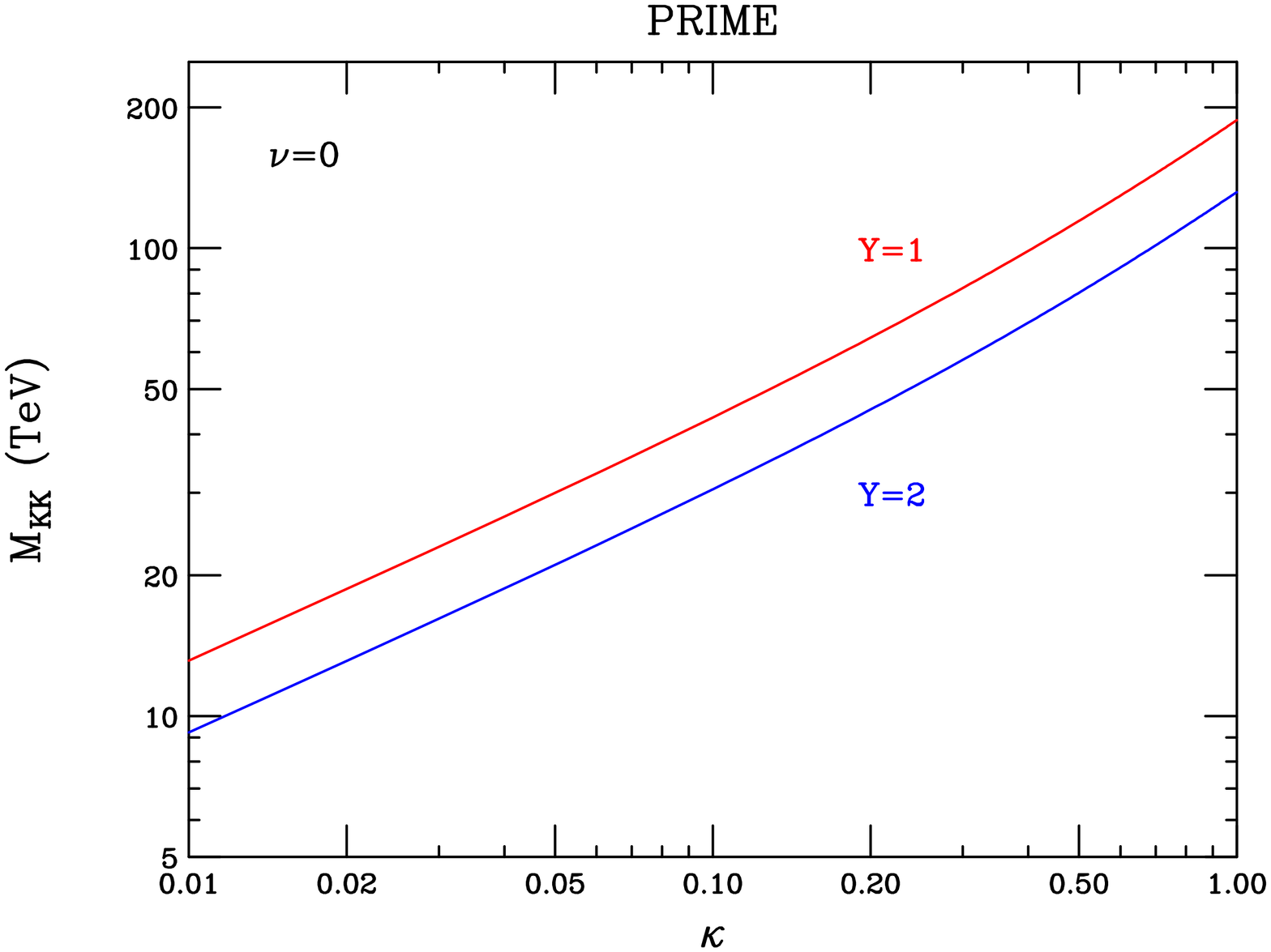,height=8.0cm,width=6.1cm,angle=90}}
\vspace{-0.2cm}
\caption{Projected bounds on $M_{KK}$ coming from MEG (left) and PRIME (right) for $\nu=0$.  We have set the mixing 
angles to $\kappa$ times their canonical values, and have varied $\kappa$ in the range $[0.01,1]$ for $Y_x=1,2$.}
\label{future}
\end{figure}

Searches for $\mu-e$ conversion and $\mu \to e\gamma$ are already starting to require slight tunings of the 
model parameters.  The limit on $BR(\mu \to e\gamma)$ is projected to improve from $1.2 \times 10^{-11}$ to $10^{-13}$ 
after MEG, while the constraint on $\mu-e$ conversion is projected to improve to $10^{-18}$ after PRIME.  The bounds 
on $M_{KK}$ that these constraints lead to are shown in Fig.~\ref{future}.  We have plotted the projected bounds as 
a function of the overall scale of the mixing angles; we have set $U^{L,R}_{12}=\kappa \sqrt{m_e/m_{\mu}}$, 
$U^{L,R}_{13}=\kappa \sqrt{m_e/m_{\tau}}$, etc., and have varied $\kappa$ in the range [0.01,1].  This tests 
how far from the natural parameters these experiments will probe.  We observe that MEG will probe $M_{KK} \leq 5$ TeV 
down to mixing angles $1/10$ times their natural sizes.  PRIME will test $M_{KK} \leq 20$ TeV down to mixing angles 
$1/10$ times their natural sizes, and will probe $M_{KK} \leq 10$ TeV down to mixing angles
$1/100$ times their canonical values.  Together, these experiments will definitively test the anarchic RS 
explanation of the flavor sector.

\section{Summary and Conclusions}

In this paper, we have studied lepton flavor violation with the SM
propagating in a warped extra dimension.  The principal motivation for this model is a solution
to the Planck-electroweak hierarchy problem.  Interestingly, there is also a solution to the flavor 
hierarchy of the SM.  The large differences in the quark and lepton masses and mixing angles 
can be explained by differing profiles of SM fermions in the extra dimension, even though 
the $5D$ Yukawa coupling are of the same size without any structure.  These profiles can 
vary substantially with small changes in the $5D$ fermion masses; no large hierarchies are 
required to account for the flavor hierarchy in the SM.  Since the Higgs field is localized near the TeV brane, 
the small masses of the first and second generations are explained by their localization near
the Planck brane.

The localization of fermion fields at different points in the extra dimension leads 
to flavor violation upon rotation to the fermion mass basis.  The 
assumption of anarchic $5D$ Yukawa couplings implies that the mixing angles 
are related to the ratios of fermion masses.  We can therefore estimate the 
leptonic mixing angles without a model of neutrino masses, unlike in the SM.  The flavor violating 
couplings are proportional to the $4D$ Yukawa interactions.  Therefore there is an 
analog of the GIM mechanism in the anarchic RS picture.  However, the sensitivities 
of lepton flavor violating experiments are large, so we expect significant constraints.  Bounds 
from electroweak precision measurements currently constrain the KK scale 
to be $M_{KK} \geq 3$ TeV, approximately.

To derive the implications of lepton flavor violating measurements for the anarchic RS scenario, 
we perform a Monte Carlo scan over the natural parameter space of this model: $O(1)$ Yukawa couplings 
and $O(1)$ variations of the mixing angles around their predicted size.  We study both the case where 
the Higgs boson is localized in the TeV brane and when it is allowed to propagate in the full $5D$ spacetime.  
We study the processes $\mu \to 3e$, $\tau \to l_1 \bar{l}_2 l_3$, $\mu-e$ conversion, and dipole decays 
of the form $l \to l^{'}\gamma$.  In the brane Higgs case, cut-off effects render the dipole decays 
uncalculable in the $5D$ RS theory; this arises from the fact that the $5D$ Yukawa couplings in this 
case have mass dimension $[-1]$, and cut-off scale effects are as large as those from KK modes.  The bulk Higgs case does 
not suffer from this drawback.

We find strong constraints throughout the entire natural RS parameter space.  The minimal allowed 
KK scale is 3 TeV, and this is permitted only for a very few points in our scan.  In the bulk 
Higgs case, this occurs partially because of a tension between the tree-level mediated $\mu-e$ 
conversion process and the loop-induced decay $\mu \to e\gamma$.  These processes have opposite 
dependences on the $5D$ Yukawa couplings, making it difficult to decouple the effects of flavor violation.  
There are a couple of possible ways to avoid these constraints.  First, the KK scale can be raised slightly 
to 5 TeV, which allows large regions of the natural RS parameter space to be realized.  However, this 
increases the fine-tuning in the electroweak sector, and will make it difficult to find the KK states 
present in this model at the LHC.  Another possibility is to reduce the leptonic mixing angles slightly, 
implying some structure in the $5D$ Yukawa matrix and indicating that the observed flavor structure 
cannot be generated completely via geometry.

There are also several possible model-building possibilities to relax these constraints.  Models with 
custodial isospin based on the gauge structure $SU(2)_L \times SU(2)_R \times U(1)_{ B - L }$ contain 
an additional $Z^{'}$ and possibly additional fermions.  The coupling of the $Z^{'}$ to the SM 
fermions is model-dependent~\cite{Agashe:2006at}, and can possibly be used to cancel some of the flavor-violating 
contributions we have studied.  These models also contain an additional right-handed neutrino 
that contributes to loop-induced dipole decays.  There is no zero-mode partner of this 
right-handed neutrino, and this contribution is therefore independent of the neutrino mixing 
parameters.  Even an $O(1)$
suppression suffices to reduce the KK scale to the 3 TeV level, opening up more 
parameter space for study at the LHC.

The definitive test of whether the observed flavor structure can be explained by the anarchic RS scenario 
will come from future lepton flavor violating measurements.  $B$-factories are currently probing mixing 
in the third generation using rare $\tau$ decays.  These constraints will improve by an order of magnitude 
with data from a super-$B$ factory, probing KK scales up to 5 TeV.  These measurements probe different 
model parameters than $\mu-e$ conversion and rare $\mu$ decays, and are therefore complimentary to these other 
experiments.  Improvements in the sensitivities of $\mu \to e\gamma$ and $\mu-e$ conversion of several 
orders of magnitude will be accomplished by the future experiments MEG and PRIME, respectively.  They 
will definitively test the geometric origin of flavor structure; for example, PRIME will probe KK scales of 
$M_{KK} \geq 10$ TeV down to model parameters $1/100$ of their natural size.  These experiments will either 
confirm or completely invalidate this geometric origin of flavor.

In conclusion, the anarchic RS picture is an attractive solution to both the electroweak 
and flavor hierarchies in the SM.  Measurements at the LHC, at future $B$-factories, and with the experiments 
MEG and PRIME will determine whether it is indeed realized in nature.

\bigskip\bigskip

\noindent
{\Large \bf Acknowledgements}

\bigskip

\noindent
We thank R. Contino, H. Davoudiasl, D. E. Kaplan,
R.~Kitano and R. Sundrum for useful discussions.  We thank A. de Gouv\^ea for catching a 
numerical error in the $\tau \to 3l$ decay modes in a first version of this paper.  
K. A. is supported in part by 
DOE grant DE-FG02-90ER40542.
A. B. is partially supported by the U.S. National Science Foundation under grant PHY-0401513.  
F. P. is supported in part by the University of Wisconsin 
Research Committee with funds provided by the Wisconsin Alumni Research Foundation.  

\appendix
\bigskip\bigskip
\noindent
{\bf APPENDIX}
\bigskip

In this Appendix we present the expressions that appear in the fermion mass and Yukawa coupling 
matrices.  We focus on the case of a bulk Higgs field; the brane Higgs mass matrix can be obtained 
by taking the appropriate limits, as discussed below.

The fermion mass matrix  for the bulk Higgs scenario is given by 
\begin{equation}
\mathcal{M}=\left(\begin{array}{ccc}
M_D                         & \frac{v}{\sqrt{2}}\Delta_R  & 0           \\
\frac{v}{\sqrt{2}}\Delta_L  & \Delta_1                    & M_{KK}      \\
0                           & M_{KK}                      & \Delta_2
\end{array}\right).
\label{flavorbasis}
\end{equation}
$M_D$ is a $3\times 3$ diagonal matrix containing the masses of the zero-mode leptons, and 
$M_{KK}$ is the diagonal matrix with the KK masses before mixing.  We assume 
that it is proportional to the identity matrix, as the deviation from this limit is small. 
The other entries can be expressed using the following two overlap integrals:
\bea
F^{ij}&=&\frac{\int_{-\pi}^\pi d\phi\chi_H(\phi) f_L^0(\phi;c_i)f_L^1(\phi;c_j)}
        {\int_{-\pi}^\pi d\phi\chi_H(\phi) f_L^0(\phi;c_i)f_L^0(\phi;c_j)}, \label{F}\\
G_{L,R}^{ij}&=&\frac{\int_{-\pi}^\pi d\phi\chi_H(\phi) f_{L,R}^1(\phi;c_i)f_{L,R}^1(\phi;c_j)}
        {\int_{-\pi}^\pi d\phi\chi_H(\phi) f_L^0(\phi;c_i)f_L^0(\phi;c_j)}.\label{G}
\eea
$\chi_H$ is the Higgs vev profile given in Eq.~\ref{prof2}, while the
$f^n(\phi;c_i)$ can be found in~\cite{DHRoffwall}.  The remaining matrices 
in Eq.~\ref{flavorbasis} are
\bea
\frac{v}{\sqrt{2}}\Delta_R^{ij} &=& U_{Lik}(U_L^{\dag}M_DU_R)^{kj}
\times F^{kj}, \\
\frac{v}{\sqrt{2}}\Delta_L^{ij} &=& (U_L^{\dag}M_DU_R)^{ik}U_{Rkj}^{\dag}
\times F^{ki}, \\
\Delta_1^{ij} &=& (U_L^{\dagger}M_DU_R)^{ij}\times G_L^{ij}, \\
\Delta_2^{ij} &=& (U_L^{\dagger}M_DU_R)^{ij}\times G_R^{ij},
\eea
where there is no sum over the indices $i,j$ but there is over the index $k$.  

To see how this reduces to the brane Higgs case, we replace the Higgs
wavefunction with a delta function on the TeV brane.   This
sets $G_R=0$ via the boundary conditions $f^1_R(\phi=\pi)=0$, so that
$\Delta_2=0$ as in Eq.~\ref{mass}.  Also, $F^{ij}= F^j$, since the
$i$ flavor cancels out of the ratio in Eq.~\ref{F}, and
$G_L^{ij}=F^iF^j$, again matching our results for the brane Higgs.  

We now discuss the diagonalization of this matrix and the fermion Yukawa 
coupling matrix.  We first diagonalize the 
lower $2\times 2$ block containing the KK masses, to remove the mixing between 
the KK fermions.  We then diagonalize the full $3 \times 3$ matrix, to remove 
mixing between zero and KK modes.  We include only the leading $v/M_{KK}$ 
corrections.

We first consider the following simple $2\times 2$ matrix, which simulates the lower 
block of Eq.~\ref{flavorbasis}:
\begin{equation}
T=\left(\begin{array}{cc}
x & 1 \\
1 & y
\end{array}\right).
\end{equation}
We assume $x,y<1$.  This matrix is diagonalized by the following unitary transformation:
\begin{eqnarray}
V&=&\frac{1}{\sqrt{2}}\left(\begin{array}{cc}
1+\frac{x-y}{4} & 1-\frac{x-y}{4} \\
1-\frac{x-y}{4} & -(1+\frac{x-y}{4})
\end{array}\right).
\end{eqnarray}
$VTV^{\dagger}$ is diagonal with eigenvalues
$\pm(1\pm\frac{1}{2}(x+y))$ to leading order in $x$ and $y$.  
We now make this the lower $2 \times 2$ block of a diagonalization matrix $V$, identifying $X\equiv
x-y=\frac{\Delta_1-\Delta_2}{M_{KK}}$, and compute $\mathcal{M}_D=V\mathcal{M}V^{\dagger}$:
\begin{equation}
\mathcal{M}_D = \left(\begin{array}{ccc} 
M_D & \frac{v}{\sqrt{2}}\Delta_R\frac{1}{\sqrt{2}}(1+\frac{X}{4}) &
\frac{v}{\sqrt{2}}\Delta_R\frac{1}{\sqrt{2}}(1-\frac{X}{4}) \\
\frac{1}{\sqrt{2}}(1+\frac{X}{4})\frac{v}{\sqrt{2}}\Delta_L & M_{KK}+\frac{\Delta_1+\Delta_2}{2} & 0 \\
\frac{1}{\sqrt{2}}(1-\frac{X}{4})\frac{v}{\sqrt{2}}\Delta_L & 0 & -M_{KK}+\frac{\Delta_1+\Delta_2}{2} 
\end{array}\right).
\end{equation}

We now diagonalize the zero-KK mixing.  We accomplish this with the following unitary 
transformation matrix:
\begin{equation}
Y = \left(\begin{array}{ccc} 
1 & -\frac{v/\sqrt{2}}{M_{KK}}\Delta_R\frac{1}{\sqrt{2}}(1-\Gamma) &
\frac{v/\sqrt{2}}{M_{KK}}\Delta_R\frac{1}{\sqrt{2}}(1+\Gamma) \\
\frac{1}{\sqrt{2}}(1-\Gamma)\frac{v/\sqrt{2}}{M_{KK}}\Delta_L & 1 & 0 \\
-\frac{1}{\sqrt{2}}(1+\Gamma)\frac{v/\sqrt{2}}{M_{KK}}\Delta_L & 0 & 1 
\end{array} \right),
\end{equation}
where $\Gamma=\frac{\Delta_1+3\Delta_2-4M_0}{M_{KK}}$.  This removes the off-diagonal elements of the 
mass matrix.  The eigenvalues are shifted at $\mathcal{O}(v/M_{KK})$; this will be important 
for the Yukawa matrix below.  The full diagonalization matrix is $YV$.  We must also 
include the phase rotation $P=diag(1,1,-1)$ to make the eigenvalues positive.  
We can then determine the masses of the KK fermions to first order in $v/M_{KK}$:
\bea
M_{KK}^{(1)} &=& M_{KK} + \frac{\Delta_1+\Delta_2}{2}  \label{Mkk1}\\
M_{KK}^{(2)} &=& M_{KK} - \frac{\Delta_1+\Delta_2}{2}  \label{Mkk2}
\eea
These expressions for the KK masses are used when computing the
amplitude for $\mu\rightarrow e\gamma$.  The expression is valid for both the
brane and bulk Higgs scenarios.

We now determine the Yukawa coupling matrix.  
The Yukawa matrix $\Lambda$ in the flavor basis is obtained by dividing by
$\frac{v}{\sqrt{2}}$ and setting $M_{KK}=0$ in 
Eq.~\ref{flavorbasis}.  
Multiplying $YV\Lambda(YV)^{\dag}P$, we obtain the Yukawa matrix in
the mass basis
\begin{equation}
\Lambda_D = \left(\begin{array}{ccc} 
\lambda_{4D} &
\frac{1}{\sqrt{2}}\Delta_R\left[1+\left(\frac{X}{4}-\frac{\Delta_2-M_0}{M_{KK}}\right)\right]
&
-\frac{1}{\sqrt{2}}\Delta_R\left[1-\left(\frac{X}{4}-\frac{\Delta_2-M_0}{M_{KK}}\right)\right]
\\
\frac{1}{\sqrt{2}}\left[1+\left(\frac{X}{4}-\frac{\Delta_2-M_0}{M_{KK}}\right)\right]\Delta_L
      & \cdots & \cdots \\
\frac{1}{\sqrt{2}}\left[1-\left(\frac{X}{4}-\frac{\Delta_2-M_0}{M_{KK}}\right)\right]\Delta_L
      & \cdots & \cdots
\end{array}\right).
\end{equation}
We do not include the lower $2\times 2$ block since we do not need it
here.  Note that $\Delta_2\gg M_0$, so we can drop the dependence on the zero mode  
mass matrix in the off-diagonal terms.  These correspond to subleading contributions
suppressed by $1/f^2$.

\end{document}